\documentclass[12pt]{article}
\usepackage{epsfig,amsfonts,amssymb}

\usepackage{comment}
\usepackage{cite}
\usepackage[normalem]{ulem}
\usepackage[utf8]{inputenc}
\usepackage[T1]{fontenc}
\usepackage{lmodern, textcomp}
\usepackage[english]{babel}
\usepackage{csquotes,cancel}

\usepackage[left=1in,right=1in,top=.8in,bottom=.8in]{geometry}

\usepackage{eurosym, xspace}
\usepackage{graphicx, subfig, float}
\usepackage{verbatim}

\usepackage{amsmath, amsfonts, amssymb,upgreek}
\usepackage{cancel}
\usepackage{mathtools}
\usepackage{slashed}

\usepackage[usenames, dvipsnames]{xcolor}
\usepackage[amsmath, hyperref, framed]{}
\usepackage{hhline}
\usepackage{bm}
\usepackage{array, caption}
\usepackage{array,multirow}

\newcolumntype{C}[1]{>{\centering\arraybackslash}m{#1}}

\usepackage[pdftex, breaklinks=true, linktocpage,
	colorlinks=true, urlcolor=blue, linkcolor=blue, citecolor=red
]{hyperref}

\usepackage{cite}
\usepackage{tikz-feynman}
\usepackage{tikz, pgf}
\usetikzlibrary{shapes.misc}
\usetikzlibrary{snakes}
\usetikzlibrary{decorations.pathmorphing}
\usetikzlibrary{shapes}
\usetikzlibrary{fit}

\def\ZZZ{{\hbox{ Z\kern-1.6mm Z}}}
\def\RRR{{\hbox{ R\kern-2.4mm R}}}
\def\CCC{{\hbox{ C\kern-2.0mm C}}}
\def\zzz{{\hbox{z\kern-1mm z}}}

\newcommand{\qeq}{{\hbox{=\kern-2.3mm ? \kern.5mm }}}
\renewcommand{\qeq}{=}

\newcommand{\be}{\begin{eqnarray}}
\newcommand{\ee}{\end{eqnarray}}

\newcommand{\vp}{\varphi}

\newcommand{\ben}{\begin{eqnarray}\displaystyle}
\newcommand{\een}{\end{eqnarray}}

\newcommand{\p}{\partial}

\def\one{{\hbox{ 1\kern-.8mm l}}}
\def\zero{{\hbox{ 0\kern-1.5mm 0}}}

\newcommand{\bea}[1]{\begin{eqnarray}\label{#1} }
\newcommand{\eea}{\end{eqnarray}}

\usepackage{bm}

\newcommand\non{\nonumber}

\newcommand\f{\frac}

\def\figone{

\def\JPicScale{0.8}
\ifx\JPicScale\undefined\def\JPicScale{1}\fi
\unitlength \JPicScale mm
\begin{picture}(135,80)(0,0)
\linethickness{0.3mm}
\multiput(40,80)(0.12,-0.18){167}{\line(0,-1){0.18}}
\linethickness{0.3mm}
\multiput(30,70)(0.18,-0.12){167}{\line(1,0){0.18}}
\linethickness{0.3mm}
\put(30,50){\line(1,0){30}}
\linethickness{0.3mm}
\multiput(30,30)(0.18,0.12){167}{\line(1,0){0.18}}
\linethickness{0.3mm}
\multiput(40,20)(0.12,0.18){167}{\line(0,1){0.18}}
\linethickness{0.3mm}
\put(60,50){\line(1,0){40}}
\linethickness{0.3mm}
\multiput(100,50)(0.12,0.18){167}{\line(0,1){0.18}}
\linethickness{0.3mm}
\multiput(100,50)(0.18,0.12){167}{\line(1,0){0.18}}
\linethickness{0.3mm}
\put(100,50){\line(1,0){30}}
\linethickness{0.3mm}
\multiput(100,50)(0.18,-0.12){167}{\line(1,0){0.18}}
\linethickness{0.3mm}
\multiput(100,50)(0.12,-0.18){167}{\line(0,-1){0.18}}

\put(30,80){\makebox(0,0)[cc]{$\zeta Q_B A_1^c$}}

\put(25,70){\makebox(0,0)[cc]{$A_2^c$}}

\put(25,50){\makebox(0,0)[cc]{$A_n^c$}}

\put(20,30){\makebox(0,0)[cc]{$A_1^o$}}

\put(35,20){\makebox(0,0)[cc]{$A_p^o$}}

\put(35,60){\makebox(0,0)[cc]{$\vdots$}}

\put(40,30){\makebox(0,0)[cc]{$\vdots$}}

\put(125,80){\makebox(0,0)[cc]{$B_1^c$}}

\put(135,70){\makebox(0,0)[cc]{$B_2^c$}}

\put(135,50){\makebox(0,0)[cc]{$B_m^c$}}

\put(135,30){\makebox(0,0)[cc]{$B_1^o$}}

\put(120,15){\makebox(0,0)[cc]{$B_q^o$}}

\put(115,35){\makebox(0,0)[cc]{$\vdots$}}

\put(120,57){\makebox(0,0)[cc]{$\vdots$}}

\put(66,47){\makebox(0,0)[cc]{$\psi_r^c\vp_r$}}

\put(93,47){\makebox(0,0)[cc]{$\psi_s^c\vp_s$}}

\end{picture}

}

\def\figtwo{

\def\JPicScale{0.8}
\ifx\JPicScale\undefined\def\JPicScale{1}\fi
\unitlength \JPicScale mm
\begin{picture}(135,80)(0,0)
\linethickness{0.3mm}
\multiput(40,80)(0.12,-0.18){167}{\line(0,-1){0.18}}
\linethickness{0.3mm}
\multiput(30,70)(0.18,-0.12){167}{\line(1,0){0.18}}
\linethickness{0.3mm}
\put(30,50){\line(1,0){30}}
\linethickness{0.3mm}
\multiput(30,30)(0.18,0.12){167}{\line(1,0){0.18}}
\linethickness{0.3mm}
\multiput(40,20)(0.12,0.18){167}{\line(0,1){0.18}}
\linethickness{0.3mm}
\put(60,50){\line(1,0){40}}

\put(30,80){\makebox(0,0)[cc]{$\zeta Q_B A_1^c$}}

\put(25,70){\makebox(0,0)[cc]{$A_2^c$}}

\put(25,50){\makebox(0,0)[cc]{$B_m^c$}}

\put(20,30){\makebox(0,0)[cc]{$A_1^o$}}

\put(35,20){\makebox(0,0)[cc]{$B_q^o$}}

\put(35,60){\makebox(0,0)[cc]{$\vdots$}}

\put(40,30){\makebox(0,0)[cc]{$\vdots$}}

\put(66,47){\makebox(0,0)[cc]{$\psi_r^c\vp_r$}}

\put(93,45){\makebox(0,0)[cc]{$\tilde\psi_s^c\tilde\vp^s$}}

\put(100,50){\makebox(0,0)[cc]{$\times$}}

\end{picture}

}

\definecolor{armygreen}{rgb}{0.29, 0.33, 0.13}

\begin{document}

\baselineskip 24pt

\begin{center}
{\Large \bf Soft Theorems from Compactification  }

\end{center}

\vskip .6cm
\medskip

\vspace*{4.0ex}

\baselineskip=18pt

\begin{center}

{\large 
\rm Raffaele Marotta$^a$ and Mritunjay Verma$^a$ }

\end{center}

\vspace*{4.0ex}

\centerline{ \it \small $^a$Istituto Nazionale di Fisica Nucleare (INFN), Sezione di Napoli, }
\centerline{ \it \small Complesso Universitario di Monte S. Angelo ed. 6, via Cintia, 80126, Napoli, Italy.}

\vspace*{1.0ex}
\centerline{\small E-mail:  raffaele.marotta@na.infn.it, 
mverma@na.infn.it }

\vspace*{5.0ex}

\centerline{\bf Abstract} \bigskip

We analyze the single subleading soft graviton theorem in $(d+1)$ dimensions under compactification on $S^1$. This produces the single soft theorems for the graviton, vector and scalar fields in $d$ dimension. For the compactification of $11$-dimensional supergravity theory, this gives the soft factorization properties of the single graviton, dilaton and RR 1-form fields in type IIA string theory in ten dimensions. For the case of the soft vector field, we also explicitly check the result obtained from compactification by computing the amplitudes with external massive spin two and massless finite energy states interacting with soft vector field. 
The former are the Kaluza-Klein excitations of the $d+1$ dimensional metric. Describing the interaction of the KK-modes with the vector field at each level by the minimally coupled  Fierz-Pauli Lagrangian, we find agreement with the results obtained from the compactification if the gyromagnetic ratio in the minimally coupled Fierz-Pauli Lagrangian is taken to be $g=1$. 
\vfill

\vfill \eject

\baselineskip18pt

\tableofcontents

\section{Introduction } 
\label{s1}
During the past few years, the soft theorems have been investigated in a variety of theories \cite{1103.2981,1404.4091,1404.7749,1405.1410,1405.2346,1405.1015,1405.3413,1405.3533,1406.6574,
1406.6987,1406.7184,1407.5936,1407.5982,1408.4179,1410.6406,1412.3699,1504.01364,1406.4172,1406.5155,1411.6661,1502.05258,
1505.05854,1507.08829,1511.04921,1512.00803,1601.03457,1604.03355,1610.03481,1611.07534,
1612.05886,ademollo,shapiro,1503.04816,1504.05558,1504.05559,1507.00938,1604.02834,1607.02700,1702.02350,1705.06175,Chakrabarti:2017zmh,DiVecchia:2018dob,DiVecchia:2019kle,Laddha:2018rle,Laddha:2019yaj} together with their connection with the asymptotic symmetries and the memory effects \cite{1312.2229,1401.7026,1411.5745,1506.05789,1509.01406,1605.09094,1608.00685,1612.08294,1701.00496}. In a series of works \cite{1702.03934,1703.00024,1706.00759,Chakrabarti:2017ltl, AtulBhatkar:2018kfi}, it was shown that the subleading soft graviton theorem is universal in higher than four space-time dimensions at any loop order.\footnote{ {At tree level,  universality holds  also for 4 dimensional space-times. In 5 dimensions, the individual Feynman diagrams at loop level can suffer from the IR divergences \cite{Chakrabarti:2017ltl}. However, they are expected to cancel in the full amplitude \cite{d=5, d=51}}.} Any theory possessing diffeomorphism invariance satisfies the soft graviton factorization up to subleading order with results valid at all loops and the form of the soft operators up to subleading order takes the same form in all these theories. At subsubleading order, the soft factorization still happens though the form of the soft operator is not fully determined from the diffeomorphism invariance alone but depends upon the theory \cite{1406.5155,1604.03355,1706.00759}. In 4 space-time dimensions, there are subtleties and the relation between amplitudes with and without soft gravitons involve logarithmic terms for loop amplitudes \cite{Laddha:2018myi ,Laddha:2018vbn, Sahoo:2018lxl}. 

\vspace*{.07in}It is also known that the leading soft photon theorem is also universal and depends only upon the gauge invariance of the theory \cite{Low,weinberg1}. At subleading order, the soft factorization still happens though the form of the soft operator depends upon the theory\cite{1611.07534}. The soft behaviour of the scalar fields have also been investigated in \cite{1512.03316,1711.05773,1810.05634}. 

\vspace*{.07in}In this paper, we focus on the single subleading soft graviton theorem. Its universality allows us to apply this to an arbitrary loop amplitude in higher than 4 space-time dimensions irrespective of the theory. We shall investigate the implications of this theorem when we compactify a space direction along a circle. The compactification of the metric field on a circle gives rise to 3 massless fields, namely, a graviton, one vector and one scalar field along with a tower of massive spin-2 fields in one lower dimension. We shall show that the soft factorization of the amplitude containing a single soft graviton in higher dimension implies the soft factorization of amplitudes involving a soft graviton, a vector and a scalar field in the lower dimension after compactification. 

\vspace*{.07in}We shall work in generic dimensions. Hence, we shall consider the soft theorem for gravitons in $d+1$ dimensions. The case of $d+1=11$ is interesting since compactifying 11 dimensional theory on $S^1$ gives the type IIA string theory in 10 dimensions \cite{Witten:1995ex, sen_m_theory}. In particular, the 11 dimensional graviton gives rise to graviton, dilaton and RR 1-form in the type IIA theory. Moreover, the massive KK modes are identified with the D0 branes and its bound states in string theory. Hence, the results for the soft factorization of vector and scalar essentially gives the soft behaviour of the dilaton and RR 1-form fields in string theory. Specializing the external finite energy states to massive KK modes in the result of compactification gives the prediction for the scattering of D0 branes and its bound states from the soft fields.

\vspace*{.07in}Before proceeding further, we summarize our results for the soft factorization of particles which arise due to compactification on $S^1$. For the graviton, the subleading result is universal. Hence, both the $d+1$ and $d$ dimensional results take the same form. For the vector field arising due to compactification on $S^1$, the subleading soft behaviour turns out to be  
\begin{eqnarray}
\!{\cal M}_{n+1}(\varepsilon,q;p_i)
=\sqrt{2}\,\kappa_d \sum_{i=1}^n \left[\frac{e_i\varepsilon_\mu p_i^\mu }{p_i\cdot q}+\frac{e_i \varepsilon_\mu\,q_\nu (2L_i^{\mu\nu}+S_i^{\mu\nu})}{2\;p_i\cdot q} +\frac{\varepsilon_\mu q_\nu p_i^\sigma (\Sigma_{\sigma \rho})^{\mu\nu}\mathcal{S}_i^{\rho }}{\sqrt{2}\;p_i\cdot q} \right]{\cal M}_n(p_i)\label{2.39intro}
\end{eqnarray}
In the above expression, the indices $\mu,\nu\cdots$ run over the $d$ dimensional space-time and take the value from $0$ to $d-1$. The $(q,\varepsilon)$ denote the momentum and polarization of the soft photon, $(p_i,\varepsilon_i)$ denote the momenta and polarization of the finite energy particles. The $e_i$ denotes the charge of the particle which interacts with the photons.\footnote{This is the charge which arises from the KK reduction.} The $L_i^{\mu\nu}$ and $S^{\mu\nu}_i$ denote the orbital and spin angular momentum respectively which act on the finite energy states inside $\mathcal{M}_n$.  The operator $(\Sigma_{\rho\sigma})_{\mu\nu}$ is defined by
\be
(\Sigma_{\sigma \rho})_{\mu\nu}&\equiv&\eta_{\sigma \mu}\,\eta_{\rho\nu}-\eta_{\sigma \nu}\,\eta_{\rho \mu}\label{spin_operator1a}.
\ee 
The $\mathcal{S}^\mu$ is an operator which annihilates massive spin-2 KK modes and acts on the massless vector $\varepsilon^\mu$, scalar $\hat\phi$ and the graviton field $\varepsilon^{\mu\nu}$ as\footnote{See also the comment below \eqref{below}.}
\begin{eqnarray}
{\cal S}^\nu\, \varepsilon^\mu&=& - \left[\varepsilon^{\nu\mu}(p_i)-\eta^{\nu\mu}\hat{\phi} \sqrt{\frac{d-1}{d-2}}\right]\qquad;\qquad{\cal S}^\mu\,\hat{\phi} =-\sqrt{\frac{d-1}{d-2}}\varepsilon^\mu   \nonumber\\[.1cm]
{\cal S}^\nu \,\varepsilon^{\rho\sigma}&=& \frac{1}{2} \Bigl[\eta^{\sigma \nu}\,\eta^{\rho\mu}+\eta^{\rho \nu}\,\eta^{\sigma\mu}-\frac{2\eta^{\mu \nu}\eta^{\rho\sigma}}{d-2}\Bigl]\varepsilon_\mu\label{1.3}
\end{eqnarray}
In other words, $\mathcal{S}^\mu$ converts one type of finite energy particle into another particle and implies the existence of specific interaction vertices. The operator $\cal{S}^\mu$ defined in equation \eqref{1.3} is essentially the broken generator, $S_i^{\rho z}$ of the spin angular momentum of $d+1$ dimensions.\footnote{We parametrize the compact direction $S^1$ by $z$ variable.} This gives an interesting connection between soft theorems and symmetries \cite{1406.6574,1406.6987, 1605.08697,1705.10078,1802.05999} which are eventually broken as in the case of spontaneously broken conformal field theories \cite{1507.08162,1705.06175}.

\vspace*{.07in}To derive the above result, we started from the universal subleading soft graviton theorem in $d+1$ dimensions and compactified it on a circle. Universality also implies the validity of the theorem for arbitrary finite energy states which are now unspecified. 
After determining the precise map between the polarizations of the $d+1$ graviton states, reduced to $d$-dimensions, and the physical polarizations of the $d$ dimensional fields, we use it in the expression of the compactified soft graviton theorem. Doing this, the subleading soft factorization properties of the graviton, vector and scalar fields in $d$ dimension are easily obtained. In this work, the above mapping will be determined only at the level of classical fields.\footnote{We thank Ashoke Sen for discussions regarding this point.} This implies that the soft graviton factors beyond subleading order and the soft photon factors beyond the leading order will be valid only at the tree level. The loop corrections might change the form of these soft operators. The subleading soft graviton operator and the leading soft photon operator are fixed universally by symmetry arguments and hence they can't change by loop corrections. All the soft scalar operators can also change by the loop corrections.  

\vspace*{.07in}We also explicitly check the consistency of the result obtained in \eqref{2.39intro} by some direct calculations.  In particular, we consider all possible exchange diagrams with the soft vector interacting with on-shell finite energy massless vectors, gravitons, scalars as well as massive spin two particles. The action for these fields can be obtained by compactifying the $d+1$ dimensional Einstein-Hilbert action keeping also the infinite tower of massive Kaluza-Klein states. The action of the massive KK states, neglecting the interaction between different KK levels and the interaction with the massless gauge field, is described by the free Fierz-Pauli Lagrangian \cite{Fierz&Pauli}. The Fierz Pauli action for the KK modes using the procedure of compactification has been obtained in \cite{cho&zho}. The interaction between the KK modes with the abelian gauge field can then be introduced by the minimal coupling procedure (see for example \cite{Deser&Waldron}). This introduces an arbitrariness in the construction of the Lagrangian which is parametrized  by an unfixed constant $g$ known as gyromagnetic ratio. This constant appears in the soft operator and the result of explicit calculation matches with the result obtained through compactification if the gyromagnetic ratio is taken to be equal to 1.  This value, obtained through soft theorems,  is in agreement with previous results showing that all the Kaluza-Klein states have gyromagnetic ratio $g=1$\cite{HIOY:84,9801072}\footnote{ We thank M. J. Duff for bringing to our notice these papers regarding the gyromagnetic factor of massive Kaluza-Klein states.}. 
 
 In this approach the definition of the operator ${\cal S}_i^\rho$, given in equation \eqref{1.3} is quite formal; it converts a vector to a scalar/graviton state and vice-versa.  {Its} action reflects the peculiarity of the three point vertices where a vector field can interact only with another vector field and a scalar or graviton state. However, after comparing the result of direct calculation with the result obtained from compactification, the introduction of the operator ${\cal S}_i^\rho$ becomes quite natural being identified with the broken generator of the $d+1$ dimensional spin operator. The demand of the gauge invariance of the whole on-shell amplitude then fixes the structure of the soft operator. 
 
\vspace*{.07in}For the amplitudes involving a soft scalar field and  hard particles having spin zero, one or two, the prediction for the leading and subleading soft behaviour which follows from compactification, is given by 
\begin{eqnarray}
\mathcal{M}_{n+1} = \kappa_{d}\left\{2+\sum_{i=1}^n\left[ \frac{(2-d)e_i^2-m_{i}^2 }{p_i\cdot q}+\frac{(2-d)e_i^2-m_i^2}{p_i\cdot q} q\cdot \frac{\partial}{\partial p_i}-p_i\cdot \frac{\partial}{\partial p_i}-(2-d)e_i\frac{\partial}{\partial e_i}\right]\right\}\mathcal{M}_{n}\nonumber\\
\label{2.41intro}
\end{eqnarray}  
where
\begin{eqnarray}
m_i^2 =(m_i^{d+1})^2+  \frac{n_i^2}{R^2}\qquad;\qquad e_i=\frac{n_i}{R_d}
\end{eqnarray}
with $m_i^{d+1}$ being the mass of the finite energy state in $d+1$ dimensions, zero if massless.  The $n_i^2/R_d^2$ is the contribution coming from the compactification along $S^1$. The $R_d$ denotes the radius of the circle and $n_i$ is an integer labelling the tower of massive KK-states. The $e_i$ denote the charge of the massive spin 2 KK modes which are charged under the massless gauge field.
For uncharged particles ($n_i=0$),  equation \eqref{2.41intro} is in agreement with the previous results obtained in the literature for the string dilaton \cite{1512.03316}. This result therefore supports the universality of such a factorization property. The presence of charged massive states add  new terms to the soft scalar operator.

\vspace*{.07in}The rest of the paper is organised as follows. In section \ref{sec2:identification}, we make the connection between the polarization tensors in $d+1$ and $d$ dimensions. A crucial aspect of this is the relation between the zero mode of the graviton polarization in $d+1$ dimensions and the polarization of the massless fields in d dimensions. In section \ref{s2}, we analyze the single subleading soft graviton theorem in $d+1$ dimensions under compactification and show that it gives the soft factorization of massless graviton, vector and the scalar field in $d$ dimensions. In section \ref{s3}, we perform some explicit checks. In particular, we show  that the soft factorization of amplitudes involving a single vector field which is obtained through compactification also follows by a direct calculation. For this, we consider both massless as well as massive spin 2 finite energy external states interacting with the soft vector field. The identification of the soft generators in $d$ dimensions in terms of the angular momentum operator involving the compact direction is very crucial in this matching.
We end with some discussion in section \ref{s4}. In appendix \ref{apps1}, we review some results about the compactification of metric in $d+1$ dimensions on $S^1$.

\section{Identifying the physical polarizations}
\label{sec2:identification}
To show that the soft factorization of graviton in $d+1$ dimensions gives rise to soft factorization of graviton (and other particles resulting from the compactification) in $d$ dimensions, we first need to identify the correct physical polarizations in $d$ dimensions in terms of the graviton polarizations of $d+1$ dimensions. This is the question we shall address in this section. 

\vspace*{.07in} The gravity in $d+1$ dimensions, whose compactification to $d$-dimensions is discussed in appendix, is described by the Einstein-Hilbert action \eqref{app1e}. We shall 
parametrize metric as
\begin{eqnarray}
G_{MN}=\eta_{MN}+2\kappa_{d+1} S_{MN}\label{2.7}
\end{eqnarray}  
and introduce the $d+1$ dimensional on-shell graviton field $S_{MN}=\epsilon_{MN} e^{ip_Mx^M}$, whose polarization tensor satisfies the following conditions
\begin{eqnarray}
p_M\epsilon^{MN}=0=\epsilon^N_{~N}~~;~~\epsilon_{MN}=\epsilon_{NM}~~;~~p^2=0\label{2.6}
\end{eqnarray}
with $M,N=0\dots d$. 

\vspace*{.07in}We shall denote by $\mu,\nu=0,\cdots,d-1$ the indices along the $d$ dimensional non compact space-time and by $z$ the index along the compact direction. In this notation, the conditions on the $d+1$ dimensional polarization tensor can be written as
\be
p_\mu \epsilon^{\mu\nu}+ p_z \epsilon^{z\nu}=0\quad,\quad p_\mu \epsilon^{\mu z}+ p_z \epsilon^{zz}=0\quad,\quad \epsilon^\mu_{\;\;\mu}+\epsilon^z_{\;\;z}=0\label{cond_d+1}
\ee 

\vspace*{.07in}As reviewed in appendix \ref{apps1}, the dimensional reduction of the $d+1$ dimensional metric along a circle can be performed by the following metric ansatz 
\begin{eqnarray}
G_{\mu\nu}= e^{2\alpha \phi} g_{\mu\nu} +e^{2\beta \phi } A_\mu\,A_\nu~~;~~G_{\mu z}= e^{2\beta\phi} A_\mu~~;~~ G_{zz}= e^{2\beta \phi}\label{2.8}
\end{eqnarray}
For the moment, we consider all the fields $g_{\mu\nu},A_\mu$ and $\phi$ to depend only on the $d$ dimensional coordinates $x^\mu$,  with $\mu,\nu= 0,\dots, d-1$. This is the case if the fields have to represent the massless degrees of freedom in $d$ dimensions. The constants $\alpha$ and $\beta$ are chosen to be
\begin{eqnarray}
\alpha^2= \frac{1}{2(d-2)(d-1)}\qquad, \qquad \beta =-(d-2)\alpha\label{2.9}
\end{eqnarray}
These constants are fixed by requiring that the reduced action is  in the Einsten frame and has the dilaton kinetic term  normalized with the factor $1/4\kappa_d^2$ as reviewed in appendix \ref{apps1}. For the above ansatz, the $d$-dimensional action turns out to be
\begin{eqnarray}
S=\frac{1}{2\kappa_{d}^2} \int d^{d} x \sqrt{-g} \, \left[R-\frac{1}{2} \partial_\mu\phi \partial^\mu \phi
-\frac{1}{4}e^{\sqrt{2(d-1)/(d-2)}\,\phi} \,F_{\mu\nu}F^{\mu\nu}\right]~~~~~\label{2.10}
\end{eqnarray}
where $F_{\mu\nu}$ denotes the field strength of the 1-form field $A_\mu$. This is the $d$-dimensional action for the metric, dilaton and rank 2 field strenght (see, e.g., \cite{9412184}). When $d=10$, it is the type IIA action, restricted to these three fields, obtained by dimensional reduction of the Einstein-Hilbert term of the $d=11$ supergravity. The canonically normalized kinetic terms for the fields in action (\ref{2.10}) are obtained by absorbing the gravitational coupling constant in the fields as 
\begin{eqnarray}
g_{\mu\nu} = \eta_{\mu\nu}+ 2\kappa_{d} h_{\mu\nu}~~;~~\phi=\sqrt{2}\kappa_{d}\, \hat{\phi}~~;~~ A_\mu =\sqrt{2}\kappa_{d}\,\hat{A}_\mu\label{2.11}
\end{eqnarray}
The soft theorems are expressed in terms of the momentum space polarizations. We shall denote the d dimensional graviton polarization by $\varepsilon_{\mu\nu}$. It is demanded to satisfy
\begin{eqnarray}
p^\nu\varepsilon_{\nu\mu}=
\varepsilon^\nu_{~\nu}=0
~~;~~\varepsilon_{\mu\nu}=\varepsilon_{\nu\mu}  ~,\label{2.12}
\end{eqnarray}
and similarly for the vector and scalar fields.

\vspace*{.07in} The identification between the $d$ dimensional polarization tensors and the $d+1$ dimensional polarization tensors is taken to be
\begin{eqnarray}
\epsilon_{\mu\nu}(p^\mu) &=& \frac{\kappa_{d}}{\kappa_{d+1}} \left( \varepsilon_{\mu\nu}(p^\mu)+\frac{2\alpha }{\sqrt{2}}\hat{\phi}(p^\mu)\,\eta^\perp_{\mu\nu} \right)+  O(\kappa_{d}^2)\non\\[.3cm]
\epsilon_{\mu z}(p^\mu)&=& \frac{\kappa_{d}}{\sqrt{2}\kappa_{d+1}} \varepsilon_\mu(p^\mu)+O(\kappa_{d}^2)~~;~~\epsilon_{zz}(p^\mu)= 2\beta \frac{\kappa_{d}}{\sqrt{2}\kappa_{d+1}}\hat{\phi}(p^\mu) +O(\kappa_{d}^2)
\label{2.14}
\end{eqnarray}
 $\varepsilon_\mu$ denotes the polarization of the vector field in $d$ dimension and (on-shell)
\begin{eqnarray}
\eta_{\mu\nu}^\perp\equiv\eta_{\mu\nu} -p_\mu \bar{p}_\nu -p_\nu\bar{p}_\mu\quad;\quad p\cdot \bar{p}=1\quad ;\quad p^\mu\eta_{\mu\nu}^\perp(p)=0\qquad;\qquad \bar p^2=0
\end{eqnarray} 
where $\bar p^\mu$ is a reference null vector. 

\vspace*{.07in}The terms with higher orders in $\kappa_d$ would contain terms which are not linear in the fields. 
They would correspond to vertices with more that one soft particle and therefore they do not contribute to single soft theorems. However, they would be relevant in the case of multiple soft theorem.

\vspace*{.07in}The reason for introducing the transverse metric $\eta_{\mu\nu}^\perp$ in the relation between graviton polarizations in $d+1$ and $d$ dimensions in equation \eqref{2.14} is that at the massless level, the $d+1$ dimensional fields do not depend upon momentum $p_z$ along the compact direction. Hence, the transversality condition of the $d+1$ dimensional polarization tensor, namely, $p_M\epsilon^{MN}=0$, given in equation \eqref{2.6}, immediately gives $p_\mu\epsilon^{\mu\nu}=0$. Now, the $d$ dimensional polarization tensor $\varepsilon_{\mu\nu}$ is also demanded to satisfy the same relation, namely, $p_\mu\varepsilon^{\mu\nu}=0$. However, both these conditions are not compatible with each other if we use $\eta_{\mu\nu}$ in the identification made in \eqref{2.14}. By using $\eta_{\mu\nu}^\perp$, both the transversality conditions have been made compatible with each other. Another way to see this is to note that the trace of the first line of equation \eqref{2.14} gives (on using $\varepsilon^\mu_{\;\;\mu}=0$ and equation \eqref{cond_d+1})
\begin{eqnarray}
\epsilon^\mu_{~\mu}(p)= -\epsilon^z_{~z}(p)= 2\alpha (d-2) \frac{\kappa_{d}}{\sqrt{2}\kappa_{d+1}} \hat{\phi}(p)+  O(\kappa_{d}^2)
\end{eqnarray} 
where we have used $(\eta^\perp)^\mu_{~\mu}=d-2$. 
This identity is compatible with the expression of $\epsilon_{zz}$\footnote{Note that we raise and lower the indices using the flat metric since the graviton is the fluctuation around the flat metric.} given in the second line of equation \eqref{2.14} provided the condition $\beta=- (d-2)\alpha$ holds. But, this is exactly one of the conditions given in equation \eqref{2.9} which is required by the dimensional reduction ansatz.
 
\vspace*{.07in}The  compactification along the circle, with radius $R_{d}$, also gives rise to an infinite tower of massive Kaluza-Klein states. This can be seen by expanding the field in its fourier modes. For a generic field $\Phi(x,z)$ in $d+1$ dimensions having mass $m$, we have
\begin{eqnarray}
\Phi(x,z)=\Phi^{(0)}(x)+\sum_{n\neq 0}\Phi^{(n)}(x) e^{i p_z z} 
\end{eqnarray}
with $0\leq z \leq 2\pi R_{d}$ and $p_z=n/ R_{d}$. 

\vspace*{.07in}The non zero mode $\Phi^{(n)}(x)$ represents the $n^{th}$ level massive KK field. In the compactification, the zero modes do not change their mass while for  the non-zero modes, the mass gets shifted by the presence of the compact momentum, namely, the mass of the $n^{th}$ level KK mode is given by
\be
-p^2=m^2+\left(\f{n}{R_d}\right)^2
\ee
with $p^\mu=(p_0,\dots p_{d-1})$. 

\vspace*{.07in}When $\Phi (x,z)$ is taken to be the $d+1$ dimensional graviton field $S_{MN}(x,z)$, the non zero modes $S^{(n)}_{MN}(x,z)$ describe the massive spin 2 fields. More precisely, the $S_{\mu z}^{(n)}$ and $S^{(n)}_{z z}$ components act as goldstone bosons and are eaten by the $S_{\mu\nu}^{(n)}$ field \cite{Csaki:2004ay} to give a massive spin two particle having
\begin{eqnarray}
 \f{d}{2}(d-3)+(d-2)+1=\f{1}{2}(d-2)(d+1)
 \end{eqnarray} 
 degrees of freedom. In the previous expression, the three terms on the left side are the on-shell degrees of freedom of a massless graviton, a vector and a scalar respectively in $d$ dimensions. The right hand side represents the degrees of freedom of a massive spin 2 particle in $d$ dimensions.

\vspace*{.07in}The massive KK modes $S_{\mu\nu}^{(n)}$ are also charged with respect to the massless $U(1)$ gauge field $A_\mu$ with charge given by $e_n=n/R_d$ (see appendix \ref{apps1}).

\vspace*{.07in}For the soft particle, we need to necessarily set the component of the momentum along the compact direction to be zero, i.e., $q_z=0$, since we want it to remain massless under the compactification.

\section{Soft factorization under compactification}
\label{s2}
In this section, we consider the subleading single soft graviton theorem in $d+1$ dimension and explore its consequences when we compactify one of the spatial dimension on $S^1$. We shall make use of the identification made in previous section of the physical polarization in $d$ dimensions in terms of those in $d+1$ dimensions. We start by recalling the single soft graviton theorem in $d+1$ dimensions. 

\vspace*{.07in} It is well known that an amplitude  $M_{n+1}$ which involves a graviton carrying a soft momentum $q$, and $n$ arbitrary finite energy particles, carrying momenta $p_i\ \ (i=1,\cdots,n)$ is related to the amplitude without the soft particle, $M_n$, by the so called soft graviton theorem as
\begin{eqnarray}
\mathcal{M}_{n+1}(q;\{p_i\})
\ =\ \kappa_{d+1}\left[\hat{S}^{(-1)}+\hat{S}^{(0)}+\hat{S}^{(1)}\right] \mathcal{M}_{n}(\{p_i\})+O(q^2)\label{1soft}
\end{eqnarray}
Here $\kappa_{d+1}$ is the $d+1$ dimensional gravitational coupling constant, the  $p_i$ denote the momenta of the hard particles,  $\epsilon_{MN}$ denotes the polarization of the soft graviton and the $\hat{S}^{(m)}$ $(\mbox{for}\ m=-1,\,0,\,1)$  are the soft operators to the order $q^m$ in the soft expansions. The leading and the subleading soft operators are given by
\begin{eqnarray}
\hat{S}^{(-1)}=   \epsilon_{MN}\sum_{i=1}^n\frac{p_i^M\,p_i^N}{p_i\cdot q}\qquad,\qquad\hat{S}^{(0)}=   \epsilon_{MN}\,\sum_{i=1}^n  \frac{q_Pp_i^M\, J_i^{NP}}{p_i\cdot q} \label{2.2}
\end{eqnarray} 
with $J_i^{MN}$ denoting the total angular momentum operator acting on the polarization tensors of finite energy states inside $M_n$. It is given by the sum of orbital and spin angular momentum 
\begin{eqnarray}
J_i^{MN}= L_i^{MN}+S_i^{MN}~~ ;~~ L_i^{MN}=p_i^M\frac{\partial}{\partial p_{iN}}-p_i^N\frac{\partial}{\partial p_{iM}}~~~~
\label{2.3}
\end{eqnarray}
The spin angular momentum operator $S_i^{MN}$ takes different representations depending on what finite energy state it acts upon. E.g., its action on a spin-2 state is given by
\be \label{edefJ}
(S^{MN} \, \epsilon)_{PQ} =(S^{MN})_{PQ}^{\;\;\;\;\;AB}\epsilon_{AB}
= -\bigl(\delta^M_P \epsilon^N_{~Q} - \delta^N_P \epsilon^M_{~Q}
+\delta^M_Q \epsilon^{~N}_{P} - \delta^N_Q \epsilon^{~M}_{P}\bigl)
\, .
\ee
The leading and the subleading soft operators $\hat{S}^{(-1)}$ and $\hat{S}^{(0)}$ are universal and hence independent of the particular theory we consider. Moreover, the above soft theorem statement is valid for any kind of finite energy particles. On the other hand, the subsubleading operator $\hat{S}^{(1)}$, whose explicit form is given in \cite{1706.00759}, is not universal and depends upon the specific interactions of the theory under considerations. E.g., in the case of Heterotic and Bosonic string theory in 10 dimensions, it depends upon the interaction term $\phi R^2$, (with $\phi$ being the dilaton field) which is present in the effective actions of these theories to the $O(\alpha')$, with $\alpha'$ being the string slope \cite{1406.5155, 1610.03481}.

\vspace*{.07in}The universality of the first two terms in equation \eqref{1soft} allows us to apply the theorem to an arbitrary theory describing gravity in $d+1$ dimensions and consider the scenario in which one of the direction is compactified on $S^1$. We shall consider the leading and subleading cases separately. However, before proceeding further, we make some comments about the notation. By extracting out the polarization tensors, we express the single soft graviton amplitude ${\cal M}_{n+1}$ in $d+1$ dimensions as
\begin{eqnarray}
{\cal M}_{n+1}(q,p_1,\dots p_n)= \kappa_{d+1} \epsilon_{MN}(q) {\cal M}^{MN}_{n+1}(q,\dots p_i,\dots p_n)\label{2.4}
\end{eqnarray}
where, $\epsilon_{MN}(q^\mu,q^z)$ is the polarization of the soft particle.

After we compactify on $S^1$, the massless (and hence soft) modes do not depend upon the compact direction. Using the identification made in equation \eqref{2.14}, the soft graviton amplitude $\mathcal{M}_{n+1}$ can be expressed as the sum of three terms\footnote{The precise normalization constants appearing in the right hand side of \eqref{2.1} are somewhat arbitrary. However, they are very convenient in analyzing the soft graviton theorem under compactification. This also means that these definitions of the d-dimensional amplitudes, suggested from the 
compactification procedure, could differ by an overall normalization from  the ones obtained by 
using more conventional approaches. }
\begin{eqnarray}
{\cal M}_{n+1}(q, \epsilon,\{p_i\})
&= &\kappa_d\,\varepsilon_{\mu\nu} \mathcal{M}^{\mu\nu}_{n+1}(q, \varepsilon,\{p_i\})+
\kappa_d\varepsilon_\mu  \mathcal{M}^{\mu}_{n+1}(q, \varepsilon,\{p_i\},)\nonumber\\[.2cm]
&&+\ \sqrt{2}\ \alpha  \hat{\phi} \mathcal{M}^\phi_{n+1}(q, \varepsilon,\{p_i\})+ O(\kappa_d^2)\label{2.1}
\end{eqnarray}
where we have denoted ${\cal M}^{\mu}_{n+1}\equiv {\cal M}^{\mu z}_{n+1}$ and defined 
\begin{eqnarray}
\mathcal{M}^\phi_{n+1}= \,\kappa_d\ \eta^\perp_{\mu\nu} 
\left(\mathcal{ M}^{\mu\nu}_{n+1}(q, \varepsilon,\{p_i\})-\eta^{\mu\nu}  \mathcal{M}^{zz}_{n+1}(q, \varepsilon,\{p_i\}) \right)
\end{eqnarray}
The $\mathcal M_{n+1}$ in the right hand side in equation \eqref{2.1} depend upon the finite energy massless and the massive states. All the states (massive as well as massless) depend upon the momentum along the $d$ non compact directions. The massive states also depend upon the compact direction through their mass/charge.

\subsection{Leading term}

By replacing the $d+1$ dimensional graviton polarization $\epsilon_{MN}$ in terms of the $d$ dimensional polarizations using equation \eqref{2.14}, we find that the leading soft theorem of equation \eqref{1soft} takes the form of the right hand side of equation \eqref{2.1}, thus, breaking into soft factorizations of three particles. For the graviton, we get\footnote{Note that the sum over indices in the inner products $(p_i\cdot q)$ in \eqref{softg1}-\eqref{2.24} runs only over the $d$ dimensional space-time unlike the inner products in \eqref{1soft} in which it runs over $d+1$ dimensional space-time. We shall use the same notation in both cases since the difference between the two is clear from the context.} 
\begin{eqnarray}
{\cal M}_{n+1}^g(q,\{p_i\})\equiv \kappa_d\,\varepsilon_{\mu\nu}{\cal M}^{\mu\nu}_{n+1}=
\kappa_{d}\sum_{i=1}^n \f{  \varepsilon_{\mu\nu} p_i^\mu p_i^\nu}{(p_i\cdot q)}
\mathcal{M}_{n}(\{p_i\} ),
\label{softg1}
\end{eqnarray}
 for the scalar, we get
\begin{eqnarray}
{\cal M}_{n+1}^\phi(q,\{p_i\})=\kappa_{d}\, \left\{2+\sum_{i=1}^n\left[\frac{(2-d)(p_{i}^z)^2 +p_i^2}{(p_i\cdot q)}\right]\right\} \mathcal{M}_{n}(\{p_i\} ),
\label{2.23}
\end{eqnarray}
By projecting the amplitude on the soft dilaton particle and using the momentum conservation, we get in Eq.\eqref{2.23} a subleading contribution that should be neglected at $O(q^{-1})$. Since in the next subsection we shall add to this result the subleading contribution, we keep also such a term in \eqref{2.23} to obtain the full $O(q^0)$ soft dilaton theorem.

\vspace*{.07in}Finally, for the vector field, we get
\begin{eqnarray}
{\cal M}^A_{n+1}(q,\{p_i\})\equiv\kappa_d\,\varepsilon_\mu {\cal M}^{\mu}_{n+1}=\sqrt{2}\, \kappa_d\, \sum_{i=1}^n \frac{e_i\varepsilon_\mu p_i^\mu }{(p_i\cdot q)}\mathcal{M}_{n}(\{p_i\} )
\qquad;\quad e_i=p_i^z\label{2.24}
\end{eqnarray}
For the single soft graviton, \eqref{softg1} is the standard leading soft graviton theorem statement in $d$ dimensions. In \eqref{2.24}, the $e_i$ is equal to $p^z_i$. As reviewed in appendix \ref{apps1}, the massive KK modes are charged with respect to the massless vector field and hence $p_i^z$ is the charge vector fields interact with. Thus, \eqref{2.24} has the form of the leading soft-theorem statement for a vector field in $d$ dimensions. This is expected since the leading soft photon theorem is universal and hence independent of the theory.

\vspace*{.07in}In the case $d=10$, the vector field corresponds to the RR 1-form field and $p_i^z$ represents the charge of the D0 brane and its bound states.  
The value of the charge, namely, $p_i^z= \frac{n}{g_s \sqrt{\alpha'}}$, where $1/(g_s \sqrt{\alpha'})$ is the charge of the single D0 brane, and equation \eqref{2.24} are
consistent with the identification of the D0 brane and its bound states as Kaluza-Klein  modes of the $d=11$  supergravity theory. 

\vspace*{.07in}The soft factorization of the scalar field in \eqref{2.23} is consistent with the corresponding result obtained for the dilaton in \cite{1512.03316}.

\subsection{Subleading term}
Again, by replacing the $d+1$ dimensional graviton polarization in terms of the $d$ dimensional polarizations in the subleading term involving $\hat{S}^{(0)}$ of \eqref{1soft}, we find
\be
&&\hspace*{-.35in}\mathcal{M}_{n+1}(\epsilon, q; \{p_i\})\non\\
&=& \kappa_{d}\sum_{i=1}^n \biggl[ \f{  \varepsilon_{\mu\nu} p_i^\mu q_\rho J_i^{\nu\rho}}{(p_i\cdot q)}+\f{\varepsilon_{\mu }q_\rho ( p_i^\mu J_i^{ z\rho}+p_i^zJ_i^{\mu\rho})}{\sqrt{2}(p_i\cdot q)}+ \f{2(\beta p_i^z q_\rho {J}_i^{ z\rho}+\alpha\eta^\perp_{\mu\nu} p_i^\mu q_\rho J_i^{\nu\rho})\hat\phi }{\sqrt{2}(p_i\cdot q)}\biggl]\mathcal{M}_{n}(\{p_i\} )\non\\ 
\label{dilpart}
\ee
We first focus on the case when the finite energy states are all massless. In this case, $p_i^z=0$ and the expression \eqref{dilpart} simplifies to give the following relations among  amplitudes with and without soft-particles
\be
\mathcal{M}^g_{n+1}&=& \kappa_{d}\sum_{i=1}^n \f{  \varepsilon_{\mu\nu} p_i^\mu q_\rho J_i^{\nu\rho}}{(p_i\cdot q)}\,\mathcal{M}_{n}\qquad;\qquad \mathcal{M}^\phi_{n+1}= \kappa_{d}\sum_{i=1}^n \f{\eta^\perp_{\mu\nu} p_i^\mu q_\rho J_i^{\nu\rho}}{(p_i\cdot q)}\,\mathcal{M}_{n}\label{gphisoft}\\[.3cm]
\mathcal{M}^A_{n+1}&=&\sqrt{2}\,\kappa_d\sum_{i=1}^n\f{ \varepsilon_{\mu }q_\rho \,p_i^\mu J_i^{ z\rho}} {2 (p_i\cdot q)} \,\mathcal{M}_{n}\ =\sqrt{2}\, \kappa_d\,\, \sum_{i=1}^n\frac{ \varepsilon^\mu q^\nu\,p_i^\sigma\,(\Sigma_{\sigma \rho})_{\mu\nu}\,J_i^{z\rho }}{2\,(p_i\cdot q)}\, {\cal M}_n
\label{dilpart12}
\ee
In going to the second equality of \eqref{dilpart12}, we have used the conservation of the angular momentum $\sum_{i=1}^n J_i^{\rho z}  {\cal M}_n=0$ and made use of the operator 
$(\Sigma_{\sigma \rho})_{\mu\nu}$ defined in equation \eqref{spin_operator1a}.

The first equation in \eqref{gphisoft} is the usual subleading term of the single soft graviton theorem in $d$ dimensions. The second equation of \eqref{gphisoft} shows the soft scalar factorization. We shall simplify this expression now. It is easy to see that the $\eta^\perp_{\mu\nu}$ in this can be replaced by $\eta_{\mu\nu}$ since the difference between the two vanishes after using the angular momentum conservation $\sum_{i=1}^nJ_i^{\nu\rho} {\cal M}_n=0$.  
Now, using equation \eqref{2.3}, we can recast the resulting expression in the form
\begin{eqnarray}
\sum_{i=1}^n \f{p_{i\mu} q_\rho J_i^{\mu\rho} }{(p_i\cdot q)}\mathcal M_n&=&\sum_{i=1}^n \left[ -p_i\cdot \frac{\partial}{\partial p_i}+\frac{p_i^2}{p_i\cdot  q} q\cdot \frac{\partial }{\partial p_i}+ \f{p_{i\mu} q_\rho S_i^{\mu\rho} }{(p_i\cdot q)}\right]\mathcal M_n\label{2.35}
\end{eqnarray}
The term proportional to $p_i^2$ is zero for the massles finite energy states. The term proportional to the spin part also vanishes for the hard massless gravitons, scalars or vector fields.\footnote{Actually, it also vanishes for the hard massive spin 2 states by following the same argument as in equation \eqref{hard_arg}.} For the scalars, it vanishes because the action of the spin operator on a scalar field is trivial. For the gravitons, it vanishes because 
\be
p_{i\mu}S_{i}^{\mu\rho} \varepsilon^i_{\nu\tau}\mathcal M_n^{\nu\tau}&=& \Bigl[p_{i\mu}\delta^\rho_{\;\nu} \varepsilon^\mu_{i\;\tau}-p_{i\nu}\varepsilon^\rho_{i\;\tau}+p_{i\mu}\delta^\rho_{\;\tau} \varepsilon^\mu_{i\;\nu}-p_{i\tau} \varepsilon^\rho_{i\;\nu}\Bigl]\mathcal M_n^{\nu\tau}\ =\ 0\label{hard_arg},
\ee    
where the second equality follows because $p_{i\nu}\mathcal{M}_{n}^{\nu\tau}=0=p_{i\tau}\mathcal{M}_{n}^{\nu\tau}$ and $p_{i\sigma}\varepsilon_i^{\sigma\tau}=0$. For the hard massless spin 1 states, on the other hand, this vanishes because
\be
p_{i\mu}S_{i}^{\mu\rho} \varepsilon^i_{\nu}\mathcal M_n^{\nu}&=& -\Bigl[p_{i\mu}\delta^\rho_{\;\nu} \varepsilon_i^\mu-p_{i\nu}\varepsilon^\rho_i\Bigl]\mathcal M_n^{\nu}
\ =\ 0
\ee    
where, again the second equality follows because $p_{i\nu}\mathcal{M}_{n}^{\nu}=0=p_{i\mu}\varepsilon_i^{\mu}$.

\vspace*{.07in}Thus, combining \eqref{2.35} with the $O(q^0)$ term of equations \eqref{2.23}, we get the subleading scalar soft factorization for the external massless hard particles to be
\be
\mathcal{M}_{n+1}^\phi(\epsilon, q;p_1,\cdots,p_n)
=\kappa_{d}\left[2- \sum_{i=1}^np_i\cdot \frac{\partial}{\partial p_i}\right] \mathcal{M}_{n}^\phi(p_1,\cdots,p_n)
\label{2.2.30}
\ee
This expression  is in agreement with the results given in the literature and obtained by computing string and field theory amplitudes \cite{1512.03316}.   

\vspace*{.07in}Finally, we consider \eqref{dilpart12} which gives the soft factorization of the vector field $\varepsilon_\mu$ to the subleading order. It is easy to see that the soft operator appearing in this equation is invariant under the gauge transformation $\varepsilon_\mu\rightarrow \varepsilon_\mu+q_\mu$. However, it contains the generators $J_i^{z\rho}$ of the Lorentz group in $d+1$ dimensions which are broken in the compactification procedure. These generators are defined in equation \eqref{2.3} in terms of the polarizations and momenta of $d+1$ dimensions. It is very instructive to rewrite this subleading vector soft operator in terms of the $d$-dimensional degrees of freedom which we do now.  

\vspace*{.07in}The finite energy massless particles do not depend upon the momentum along the compact direction. Hence, the orbital part of $J_i^{z\rho}$ doesn't play any role and we only need to consider the action of the spin operator $S_i^{z\rho }$  on the $d$-dimensional massless fields. Using the representation \eqref{edefJ} and equation \eqref{2.14}, we find the action of the spin operator on the scalar field to be 
\begin{eqnarray}
(S_i^{\rho z } \epsilon_i)_{zz}=2\epsilon_i^{\rho z} \qquad\implies\quad S_i^{\rho z } \hat{\phi}_i=\frac{1}{\beta}\varepsilon_i^\rho=\sqrt{\f{2(d-1)}{d-2}\varepsilon_i^\rho} \label{2.29ad}
\end{eqnarray}
Next, we consider the action on the spin 2 massless fields. Using the gauge conditions  $p^i_{\mu_i}{\cal M}_n^{\mu_i\nu_i}=0=p^i_{\nu_i}M_n^{\mu_i\nu_i}$ and the equations \eqref{2.14}, \eqref{edefJ} and \eqref{2.29ad}, we find 
\begin{eqnarray}
S_i^{\rho z} \bigl[ \epsilon^i_{\mu \nu} {\cal M}_n^{\mu \nu}\bigl]
&=&-\epsilon_i^{z\gamma } \left[\eta_{\gamma\nu}\,\eta^{\rho}_\mu+  \eta^{\rho}_\nu\,\eta_{\gamma\mu}\right]
{\cal M}_n^{\mu\nu}\non\\[.3cm]
\implies\qquad S_i^{\rho z} \bigl[ \varepsilon^i_{\mu \nu} {\cal M}_n^{\mu \nu}\bigl]&=&-\frac{1}{\sqrt{2}}\varepsilon_i^\sigma \left[\eta_{\nu\sigma}\,\eta^{\rho}_\mu+\eta^{\rho}_\nu\,\eta_{\sigma\mu}-\frac{2\,\eta_{\sigma \rho}\eta_{\mu\nu}}{(d-2)}\right]{\cal M}_n^{\mu\nu}
\label{2.28}
\end{eqnarray}
Equation \eqref{2.28} when combined with the second line of \eqref{dilpart12} shows that the action of the vector soft operator on a hard graviton leg transforms it into a vector field. This corresponds to a 3-point interaction vertex in which a soft vector field is attached to a hard graviton which interact with an $n$-point amplitude through the exchange of a vector field (see figure \ref{3tree}). Equation \eqref{2.29ad}, instead, corresponds to interaction among two on-shell vector fields and an internal  scalar. 

\vspace*{.07in}Finally, the action of the spin operator on massless vector fields is obtained using equations  \eqref{2.14} and \eqref{edefJ} to be
\begin{eqnarray}
S_i^{\rho z}\bigl[ \epsilon^i_{\mu z} {\cal M}_n^{\mu z}\bigl]
&=&(\epsilon^\rho_{i\;\mu} -\eta^\rho_\mu \epsilon^z_{i\;z})
{\cal M}_n^{\mu z}\non\\[.3cm]
\implies\qquad S_i^{\rho z} \frac{\varepsilon^i_\mu}{\sqrt{2}}\, {\cal M}_n^{\mu}&=&\left(\varepsilon^\rho_{i\;\mu} -\sqrt{\frac{d-1}{d-2}}\eta^\rho_\mu\,\hat{\phi}_i\right){\cal M}_n^{\mu}\label{2.2.46}
\end{eqnarray}
This  corresponds to an interaction  vertex  where the soft vector field is attached on a hard vector field and they exchange with  the $n$-point amplitude a graviton/scalar internal state.

\vspace*{.07in}Next, we extend our analysis to include the massive spin two states which are the only massive states following from the compactification of the metric field. The gravity soft theorem remains unchanged, as expected. The vector and scalar soft theorems instead get an extra contribution from the terms in equation \eqref{dilpart} proportional to the momentum along the compact direction $p_i^z$. For the vector, the subleading term becomes 
\begin{eqnarray}
{\cal M}^A_{n+1}
&=&\sqrt{2}\,\kappa_d \sum_{i=1}^n \left[\frac{e_i \varepsilon_\mu\,q_\nu J_i^{\mu\nu}}{2p_i\cdot q} +\frac{\varepsilon_\mu q_\nu p_i^\sigma (\Sigma_{\sigma \rho})^{\mu\nu}S_i^{z\rho }}{2p_i\cdot q} 
+ \frac{e_i \varepsilon_\mu\, q_\nu L_i^{\mu\nu}}{2 p_i \cdot q}\right]{\cal M}_n\label{2.39}
\end{eqnarray}
where, we have used the identities
\begin{eqnarray}
q_\nu\, p_i^\sigma (\Sigma_{\sigma\rho})^{\mu\nu} p_i^\rho =0~~;~~q_\nu\, p_i^\sigma (\Sigma_{\sigma\rho})^{\mu \nu}\frac{\partial}{\partial p_{i\rho}}\ =\ -q\cdot p_i \frac{\partial}{\partial p_{i\mu}}+p_i^\mu q\cdot \frac{\partial}{\partial p_{i}}\ =\ q_\nu L_i^{\mu\nu}\label{below}
\end{eqnarray}
In the discussion above, we have specified the action of the spin operator $S_i^{z\rho}$ on the massless states which arise due to compactification. The action of such an operator on massive spin two states is trivial, namely\footnote{The amplitude calculation performed in the next section confirms this.},
\begin{eqnarray}
S^{z\rho}_i \phi^{(n)}_{i\,\tau \sigma}=0 \label{Sonphi}
\end{eqnarray}
This is due to the fact that the operator $S^{z\rho}_i$ acting on the polarization, $\phi^{(n)}_{i\,\tau \sigma}$, of the generic $n^{th}$  level Kaluza-Klein tower, gives terms proportional to $S^{(n)}_{\mu z}$ and $S_{zz}^{(n)}$ which are eaten  by the spin two massive modes $S_{\mu\nu}^{(n)}$. The  amplitude calculation performed in  Sec. \ref{Kaluza}, in the framework of the Fierz-Pauli Lagrangian interacting with a gauge field,  confirms equation \eqref{Sonphi}. 
If we have a generic state in $d+1$ dimensions, the action of the spin operator $S_i^{z\rho}$, on the particles arising  in the compactification to $d$ dimensions, needs to be worked out explicitly.
Finally, it is straightforward to check that equation \eqref{2.39} is gauge invariant. 

\vspace*{.07in}In the case of the scalar field, the contribution of the term involving $p_i^z$ in equation \eqref{dilpart} modifies equation \eqref{2.2.30} as follows
\begin{eqnarray}
\mathcal{M}_{n+1}^\phi &=& \kappa_{d}\left\{ 2-\sum_{i=1}^n\left[p_i\cdot \frac{\partial}{\partial p_i}+\frac{(2-d)e_i^2- m_i^2}{p_i\cdot q} q\cdot \frac{\partial}{\partial p_i}-(2-d)e_i\frac{\partial}{\partial e_i}\right]\right\}\mathcal{M}_{n}
\label{2.41}
\end{eqnarray}
where we have used $(p_i^z)^2=-p_i^2\equiv m_i^2$ and $p_i^z=e_i$.
We have also used the fact that the action of $S_i^{\rho\mu}$ on the massive spin-2 hard states vanishes, i.e. $p_{i\mu}S_i^{\rho\mu} \phi^{(n)}_{i\,\tau \sigma}{\cal M}^{\tau \sigma}_n=0$  by following the same argument as in equation \eqref{hard_arg}. Note also that comments written below equation \eqref{below} regarding the spin operator $S_i^{z\rho}$ are applicable in this case as well.

\section{Some Explicit Checks}
\label{s3}
In the previous section, we showed that the compactification of the subleading soft graviton theorem in $d+1$ dimensions implies the soft factorization in $d$ dimensions of the graviton, vector and the scalar field which result from the dimensional reduction. For the graviton, the subleading soft theorem statement is universal and is valid in every dimensions. Hence, the result obtained from the compactification is expected.
Also the soft factorization statements of the scalar field, the dilaton, were already known for hard massless and massive but uncharged particles. Equations \eqref{2.23} and \eqref{2.41} have also extended such results to the case of finite energy fields charged with respect to the abelian gauge field. In this section, we shall focus on the consistency check for a soft vector field and show that the soft factorization statements for them also follow from the amplitudes computed in the framework of low energy theories obtained by compactifying on a circle the Einstein-Hilbert action. For $d=10$, the vector is nothing but the RR 1-form field of the type IIA string theory, and therefore, the check that we shall do in this section corresponds to the IR properties of amplitudes with  RR 1-form fields.

\subsection{Soft factorization for the 1-form field}
\label{s3.1}
The effective action describing gravitons, scalars and  a 1-form field is given in equation \eqref{dim_red}. For $d=10$, this action coincides with the low energy effective action  of type IIA string theory restricted to the graviton, dilaton and RR 1-form fields.

\vspace*{.07in}We shall be interested in the case of amplitudes involving the 1-form field interacting with the graviton or the scalar field. For this case, the 3-point vertex between the scalar  and 2 abelian vector fields is given by 
\begin{eqnarray}
	V_{\phi, A,A}^{\mu\nu}(q_1,\,q_2,\,q_3)=-2i\kappa_d\sqrt{\frac{d-1}{d-2}}\ \big[ q_2^\nu q_3^\mu -\eta^{\mu\nu}(q_2\cdot q_3)\big]  
\end{eqnarray}
with the off shell condition
\begin{eqnarray}
	q_{3\nu} V_{\phi, A,A}^{\mu\nu}(q_1,\,q_2,\,q_3)=0=q_{2\mu} V_{\phi, A,A}^{\mu\nu}(q_1,\,q_2,\,q_3)
\end{eqnarray}
\vspace*{.07in}Similarly, the 3-point vertex between graviton and 2 vector fields is given by 
\be
V^{\mu\nu,\rho,\sigma}_{hAA}(q_1,q_2,q_3)
&=&-2 i\kappa_d\bigg[q_{2}^{(\mu} (q_{3}^{\nu)} \eta^{\rho\sigma}-q_3^\rho \eta^{\nu)\sigma}) +(q_{2}\cdot q_3) \eta^{\rho(\mu}\,\eta^{\nu)\sigma}\non\\
&& \hspace*{.5in}- q_2^\sigma\,\eta^{\rho(\mu}\, q_{3}^{\nu)} -\frac{1}{2} \eta^{\mu\nu}(q_{2}q_3 \eta^{\rho\sigma} -q_3^\rho q_2^\sigma)\bigg]
\label{5.1.4}
\ee
This vertex satisfies the off-shell conditions
\begin{eqnarray}
	q_{2\rho}V^{\mu\nu,\rho,\sigma}(q_1,\,q_2,\,q_3)=q_{3\sigma}V^{\mu\nu,\rho,\sigma}(q_1,\,q_2,\,q_3)=0.
\end{eqnarray}
Next, we consider the propagators. The graviton propagator in the De Donder gauge is given by
\begin{eqnarray}
	\frac{-i}{2} \left[\eta_{\mu \rho} \eta_{\nu \sigma}+\eta_{\mu \sigma}\eta_{\nu \rho}-\frac{2}{d-2} \eta_{\mu \nu}\eta_{\rho\sigma}\right]\frac{1}{p^2}
\end{eqnarray}
and the propagators of the scalar and the $1$-form are obtained to be
\begin{eqnarray}
	G_{{\phi}}(p)=\frac{-i}{p^2}\qquad;\qquad G_{{A}}^{\mu \nu}(p)=-i\frac{\eta^{\mu \nu}}{p^2}
\end{eqnarray} 
By using these Feynman rules we shall compute all possible diagrams that contribute to an $(n+1)$ point amplitude with a soft vector field and $n$ hard particles which can be gravitons, scalars or vectors. The different diagrams that can contribute to such a scattering  process are: an  exchange  diagram where the soft vector particle is attached on another vector hard particle, exchange diagrams where the soft particle is attached on a graviton or a scalar external leg and a diagram where the soft particle is attached to the $n$ point amplitude without any pole in the soft momenta. After computing these diagrams, we shall fix the soft expansion of the amplitude by using the gauge invariance $\varepsilon_\mu\rightarrow \varepsilon_\mu +\chi\, q_\mu$.
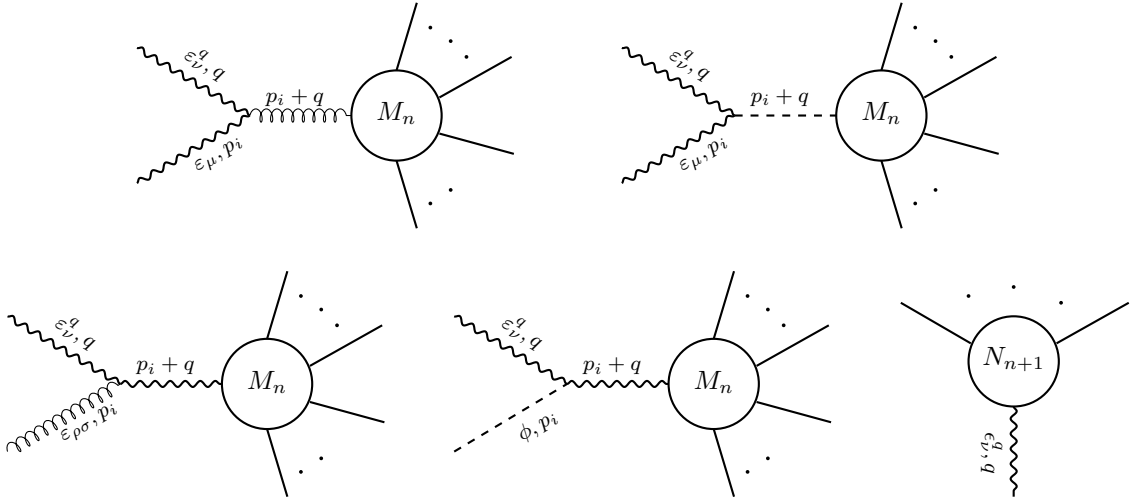
\begin{figure}
	\begin{center}
		\begin{tikzpicture}[scale=.30]
		\draw [thick]  (8.5,2) -- (9.4,5); 
		\draw [thick]  (10.4,0.8) -- (13.6,2.6); 
		\draw [thick]  (10.4,-0.8) -- (13.7,-1.7); 
		\draw [thick]  (8.5,-2) -- (9.4,-5); 
		\draw[snake=coil,segment length=4pt]         (2,0)   -- (6.5,0); 
		\draw [thick,decorate,decoration={snake,amplitude=.4mm,segment length=2mm,post length=0mm}] (-3,3) -- (2,0) node[midway, sloped,above] { \scriptsize $\hspace*{.01in}\varepsilon^q_\nu, q$};
		\draw [thick,decorate,decoration={snake,amplitude=.4mm,segment length=2mm,post length=0mm}] (-3,-3) -- (2,0) node[midway, sloped,below] { \scriptsize $\hspace*{.2in}\varepsilon_\mu, p_{i}$};
		\draw (4,.8) node { \scriptsize $ p_i+q$};
		\draw (8.5,.1) node { \footnotesize $M_n$};
		\draw [thick] (8.5,0) circle (2cm);
		\begin{scope}[shift={(-3,0)}] 
		\filldraw [ thick] (13.0,3.9) circle (1pt);
		\filldraw [ thick] (13.9,3.3) circle (1pt);
		\filldraw [ thick] (14.6,2.6) circle (1pt);
		\filldraw [ thick] (13.0,-3.9) circle (1pt);
		\filldraw [ thick] (13.9,-3.3) circle (1pt);
		\end{scope}
		\end{tikzpicture}
		\hspace*{.5in}\begin{tikzpicture}[scale=.30]
		\draw [thick]  (8.5,2) -- (9.4,5); 
		\draw [thick]  (10.4,0.8) -- (13.6,2.6); 
		\draw [thick]  (10.4,-0.8) -- (13.7,-1.7); 
		\draw [thick]  (8.5,-2) -- (9.4,-5); 
		\draw [thick,dashed]         (2,0)   -- (6.5,0); 
		\draw [thick,decorate,decoration={snake,amplitude=.4mm,segment length=2mm,post length=0mm}] (-3,3) -- (2,0) node[midway, sloped,above] { \scriptsize $\hspace*{.01in}\varepsilon^q_\nu, q$};
		\draw [thick,decorate,decoration={snake,amplitude=.4mm,segment length=2mm,post length=0mm}] (-3,-3) -- (2,0) node[midway, sloped,below] { \scriptsize $\hspace*{.2in}\varepsilon_\mu, p_{i}$};
		\draw (4,.8) node { \scriptsize $ p_i+q$};
		\draw (8.5,.1) node { \footnotesize $M_n$};
		\draw [thick] (8.5,0) circle (2cm);
		\begin{scope}[shift={(-3,0)}] 
		\filldraw [ thick] (13.0,3.9) circle (1pt);
		\filldraw [ thick] (13.9,3.3) circle (1pt);
		\filldraw [ thick] (14.6,2.6) circle (1pt);
		\filldraw [ thick] (13.0,-3.9) circle (1pt);
		\filldraw [ thick] (13.9,-3.3) circle (1pt);
		\end{scope}
		\end{tikzpicture}\\
		\vspace*{.2in}\begin{tikzpicture}[scale=.30]
		\draw [thick]  (8.5,2) -- (9.4,5); 
		\draw [thick]  (10.4,0.8) -- (13.6,2.6); 
		\draw [thick]  (10.4,-0.8) -- (13.7,-1.7); 
		\draw [thick]  (8.5,-2) -- (9.4,-5); 
		\draw [thick,decorate,decoration={snake,amplitude=.4mm,segment length=2mm,post length=0mm}]        (2,0)   -- (6.5,0); 
		\draw [thick,decorate,decoration={snake,amplitude=.4mm,segment length=2mm,post length=0mm}] (-3,3) -- (2,0) node[midway, sloped,above] { \scriptsize $\hspace*{.01in}\varepsilon^q_\nu, q$};
		\draw[snake=coil,segment length=4pt]  (-3,-3) -- (2,0) node[midway, sloped,below] { \scriptsize $\hspace*{.2in}\varepsilon_{\rho\sigma}, p_{i}$};
		\draw (4,.8) node { \scriptsize $ p_i+q$};
		\draw (8.5,.1) node { \footnotesize $M_n$};
		\draw [thick] (8.5,0) circle (2cm);
		\begin{scope}[shift={(-3,0)}] 
		\filldraw [ thick] (13.0,3.9) circle (1pt);
		\filldraw [ thick] (13.9,3.3) circle (1pt);
		\filldraw [ thick] (14.6,2.6) circle (1pt);
		\filldraw [ thick] (13.0,-3.9) circle (1pt);
		\filldraw [ thick] (13.9,-3.3) circle (1pt);
		\end{scope}
		\end{tikzpicture}
		\hspace*{.3in}\begin{tikzpicture}[scale=.30]
		\draw [thick]  (8.5,2) -- (9.4,5); 
		\draw [thick]  (10.4,0.8) -- (13.6,2.6); 
		\draw [thick]  (10.4,-0.8) -- (13.7,-1.7); 
		\draw [thick]  (8.5,-2) -- (9.4,-5); 
		\draw [thick,decorate,decoration={snake,amplitude=.4mm,segment length=2mm,post length=0mm}]        (2,0)   -- (6.5,0); 
		\draw [thick,decorate,decoration={snake,amplitude=.4mm,segment length=2mm,post length=0mm}] (-3,3) -- (2,0) node[midway, sloped,above] { \scriptsize $\hspace*{.01in}\varepsilon^q_\nu, q$};
		\draw[thick, dashed]  (-3,-3) -- (2,0) node[midway, sloped,below] { \scriptsize $\hspace*{.2in}\phi, p_{i}$};
		\draw (4,.8) node { \scriptsize $ p_i+q$};
		\draw (8.5,.1) node { \footnotesize $M_n$};
		\draw [thick] (8.5,0) circle (2cm);
		\begin{scope}[shift={(-3,0)}] 
		\filldraw [ thick] (13.0,3.9) circle (1pt);
		\filldraw [ thick] (13.9,3.3) circle (1pt);
		\filldraw [ thick] (14.6,2.6) circle (1pt);
		\filldraw [ thick] (13.0,-3.9) circle (1pt);
		\filldraw [ thick] (13.9,-3.3) circle (1pt);
		\end{scope}
		\end{tikzpicture}
		\hspace*{.3in}\begin{tikzpicture}[scale=.30]
		\draw [thick]  (6.6,0.8) -- (3.5,2.6) ;
		\draw [thick]  (10.4,0.8) -- (13.6,2.6) ;
		\draw [thick,decorate,decoration={snake,amplitude=.4mm,segment length=2mm,post length=0mm}] (8.5,-2) -- (8.5,-6) node[midway, sloped,below] { \scriptsize $\hspace*{.01in}\epsilon^q_\nu, q$};
		\draw (8.5,.1) node { \footnotesize $N_{n+1}$};
		\draw [thick] (8.5,0) circle (2cm);
		\begin{scope}[shift={(-3,0)}] 
		\filldraw [ thick] (9.5,2.9) circle (1pt);
		\filldraw [ thick] (11.5,3.3) circle (1pt);
		\filldraw [ thick] (13.6,2.6) circle (1pt);
		\end{scope}
		\end{tikzpicture}
	\end{center}
	\caption{The diagrams contributing to the calculation in section \ref{s3.1}. The 2 topmost diagrams correspond to the case when the internal state is a graviton or scalar whereas the first 2 bottom diagrams correspond to the case when the graviton or scalar are external states attached to the soft vector field. The last diagram denotes the amplitude without any pole in soft momenta. }
	\label{3tree}
\end{figure}

\vspace*{.07in}We denote the momenta of the soft vector state by $q$ and its polarization by $\varepsilon^q_\mu$ and compute each diagram contributing to the process one by one.
\begin{enumerate}
	\item The diagram with a soft vector state attached on a hard vector state and with a graviton as intermediate state is given by
	\begin{eqnarray}
		&&M_{n+1}^{(AAh)}(q,p_1,\dots p_n)\non\\
		&=& - \kappa_d \sum_{i=1}^{n_{A}}\varepsilon^q_\nu(q)\varepsilon_\mu(p_i) \Bigl[p_{i(\tau}q_{\kappa )}\eta^{\mu\nu} -p_{i(\tau} q^\mu \eta^\nu_{\kappa)}+(p_{i} \cdot q) \eta_{(\tau}^\mu\,\eta_{\kappa)}^\nu- p_i^\nu\,\eta^\mu_{(\tau}\, q_{\kappa)}\non\\
		&&-\frac{1}{2} \eta_{\tau\kappa}((p_{i}\cdot q) \eta^{\mu\nu} -q^\mu p_i^\nu)\Bigl] \left[\eta^{\tau \rho} \eta^{\kappa \sigma}+\eta^{\tau \sigma}\eta^{\kappa \rho}-\frac{2}{d-2} \eta^{\rho\sigma}\eta_{\tau \kappa}\right]\frac{1}{2q\cdot p_i}\nonumber\\
		&&M_{\rho\sigma}(p_1,\dots ,p_i+q,\dots p_n)+O(q)\nonumber\\
		&=&-2 \kappa_d \,\sum_{i=1}^{n_{A}}\f{\varepsilon^{q\nu}\varepsilon_i^{\mu}}{2p_i\cdot q}\left\{ -\frac{\eta_{\rho\sigma}}{d-2}\left[(p_i\cdot q)\eta_{\mu\nu}-p_{i\nu} q_\mu\right]   +(p_i\cdot q)\eta_{\mu(\rho}\eta_{\sigma)\nu} -p_{i\nu} \eta_{\mu(\rho}q_{\sigma)}\right\}M^{\rho\sigma}_{n}(\{p_k\})\nonumber\\ \label{pi3}
	\end{eqnarray}
	where  we have used the identity $M^{\mu\nu}=M^{\nu\mu}$ and denoted the momenta of the finite energy vector state by $p_i$. In going to the second line, we have also used 
	\begin{eqnarray}
		p_{i\mu} M^{\mu\nu}=(p_i+q)_\mu M^{\mu\nu}-q_\mu M^{\mu\nu}=-q_\mu M^{\mu\nu}
	\end{eqnarray}
	and neglected the terms of the order $O(q)$. Now, by using the expression of the spin operator given in \eqref{spin_operator1a} and by defining
	\begin{eqnarray}
		{\cal S}_\sigma \varepsilon^{\mu\nu} \equiv  \left[\eta^{(\mu}_\sigma\eta^{\nu)\rho}-\frac{\eta^\rho_\sigma\eta^{\mu\nu}}{d-2}\right]\varepsilon_\rho(p_i)
	\end{eqnarray}
	we can express \eqref{pi3} as 
	\begin{eqnarray}
		&&M_{n+1}^{(AAh)}(q,p_1,\dots p_n)=2 \kappa_d \,\sum_{i=1}^{n_{A}}\frac{\varepsilon^q_\nu p_{i\rho}q_\sigma(\Sigma^{\rho \mu})^{ \nu\sigma}{\cal S}^i_\mu}{2p_i\cdot q} M^{(h_i)}_n(p_1,\dots,p_i,\dots p_n)
	\end{eqnarray}
	where the $i$-th leg of $ M^{(h_i)}_n$ contains the graviton state.
	
	\item The diagram with a soft vector state attached on a hard vector state but with a scalar as intermediate state is given by
	\begin{eqnarray} 
		&&M_{n+1}^{(AA{\phi})}(q,p_1,\dots p_n)=- 2\kappa_d \sqrt{\frac{d-1}{d-2}} \,\sum_{i=1}^{n_{d}}\varepsilon^q_\nu\varepsilon_\mu(p_i)
		\frac{ [p_i^\nu q^\mu -\eta^{\mu\nu} (p_i\cdot q)]}{2p_i\cdot q}M_{n}(p_1,\dots ,p_i+q,\dots p_n) \nonumber\\
		&&\label{4.1.44}
	\end{eqnarray}  
	Again, by using the spin operator \eqref{spin_operator1a} and introducing an operator ${\cal S}_\mu$ such that
	\begin{eqnarray}
		{\cal S}_\mu\hat{\phi}= -\sqrt{\frac{d-1}{d-2}}\varepsilon^\nu \eta^{(\rho}_\mu\eta^{\sigma)}_\nu\eta_{\rho\sigma}=-\sqrt{\frac{d-1}{d-2}}\;\varepsilon_\mu
	\end{eqnarray}
	we can rewrite equation \eqref{4.1.44} as follows
	\begin{eqnarray}
		&&M_{n+1}^{(AAh)}(q,p_1,\dots p_n)=-2 \kappa_d \,\sum_{i=1}^{n_{d}}\frac{\varepsilon^q_\nu p_{i\rho}q_\sigma(\Sigma^{\rho \mu})^{\sigma \nu}{\cal S}^i_\mu}{2p_i\cdot q} M^{(\hat{\phi}_i)}_n(p_i,\dots,p_i,\dots p_n)
	\end{eqnarray}
	where, now the $i$-th leg of $M^{(\hat{\phi}_i)}_n$ contains a dilaton state.
	
	\item The diagrams in which the soft vector  is attached to an external graviton with another finite energy vector field as intermediate state are given by
	\begin{eqnarray} 
		&&M_{n+1}^{(Ah;A)}=-2\kappa_d\sum_{i=1}^{n_{g}}\f{\varepsilon^{q\nu}\varepsilon^{\rho\sigma}(p_i)}{2p_i\cdot q}
		\bigg[-(p_{i}\cdot q) \eta_{\mu(\rho}\,\eta_{\sigma)\nu} + p_{i\nu}\,\eta_{\mu(\rho}\, q_{\sigma)}\bigg] M_n^{\mu}(p_1,\dots ,p_i+q,\dots p_n)+O(q)\nonumber\\
		&&\label{4.1.46}
	\end{eqnarray}
	where we have used the transversality and the tracelessness conditions, namely, $p_{i\mu}\varepsilon^{\mu\nu}=0=p_{i\nu}\varepsilon^{\mu\nu}$ and $\varepsilon^\alpha_{~\alpha}=0$ respectively for the external on-shell gravitons.
	
	\item Finally, the diagram where a soft vector is attached on a finite energy external scalar leg  is given by
	\begin{eqnarray} 
		&&M_{n+1}^{(Ad;A)}= - 2\kappa_d \sqrt{\frac{d-1}{d-2}} \,\sum_{i=1}^{n_{d}}\varepsilon^{q\nu}
		\frac{ [-p_{i\nu} q_\mu +\eta_{\mu\nu} (p_i\cdot q)]}{2p_i\cdot q}M_n^{\mu}(p_1,\dots ,p_i+q,\dots p_n)\nonumber\\\label{4.1.47}
	\end{eqnarray}
\end{enumerate}
Again making use of \eqref{spin_operator1a} and introducing an operator ${\cal S}^\mu$ such that its' action on the  polarizarion vector is given by
\begin{eqnarray}
	{\cal S}^\nu \epsilon^\mu= - \left[\varepsilon_{\rho\sigma}(p_i)-\eta_{\rho\sigma}\hat{\phi} \sqrt{\frac{d-1}{d-2}}\right]\eta^{\mu(\rho}\eta^{\sigma)\nu} =- \left[\varepsilon^{\nu\mu}(p_i)-\eta^{\nu\mu}\hat{\phi} \sqrt{\frac{d-1}{d-2}}\right],
\end{eqnarray}
we can combine the contributions of the 3rd and 4th diagrams given in equations \eqref{4.1.46} and \eqref{4.1.47} to write
\begin{eqnarray} 
	&&M_{n+1}^{(Ah/d;A)}=2\kappa_d\sum_{i=1}^{n_{A}}\frac{\varepsilon^q_\nu(q) p_{i\mu}q_\rho(\Sigma^\mu_{~\sigma})^{\nu\rho} {\cal S}_i^\sigma}{2p_i\cdot q}\,M_n(p_1,\dots,p_i,\dots p_n)+O(q)\label{RR_cal}
\end{eqnarray}
The full $n+1$ point amplitude is the sum of all possible exchange diagrams and the diagram without any pole in the soft momenta
\begin{eqnarray}
	M_{n+1}^{A}(q,\,p_i,\dots ,p_n)=M_{n+1}^{(Ah/d;A)}+ M_{n+1}^{(AA;h/d)}+ N_{n+1}(q,p_1,\dots p_n)
\end{eqnarray} 
Here we have denoted $N_{n+1}=\varepsilon^q_\mu N^\mu_{n+1}$ the amplitude without any pole in the soft particle. Such contribution can be determined, in the soft region, via gauge invariance of the full amplitude, i.e.
\begin{eqnarray}
	q_\mu\,M^\mu_{n+1}(q,\,p_i,\dots ,p_n)=0\label{4.1.23}
\end{eqnarray}
It is easy to check that all the exchange diagrams  are gauge invariant  by themselves. Hence, equation \eqref{4.1.23} simply implies
\begin{eqnarray}
	q_\nu\,N^\nu(q,p_1,\dots p_n)= 0\Rightarrow N^\nu(0,p_1,\dots p_n)=0+O(q)\label{4.1.30}
\end{eqnarray}
We can now see that the contribution of all the Feynman diagrams in figure \ref{3tree} is consistent with the result obtained by dimensional reduction \eqref{2.39} when we specialize \eqref{2.39} to external massless states provided we identify $S_i^{z\rho}$ of \eqref{2.39} with the operators $\mathcal{S}_\mu$ defined above. This is an important point to note about the connection between the operators $\cal S_\mu$ and the broken generators of Lorentz group in $d+1$ dimension. Noting again the action of the operators $\cal S_\mu$ on various polarization tensors 
\begin{eqnarray}
	&&{\cal S}^\nu\, \varepsilon^\mu= - \left[\varepsilon^{\nu\mu}(p_i)-\eta^{\nu\mu}\hat{\phi} \sqrt{\frac{d-1}{d-2}}\right]\qquad;\qquad{\cal S}^\mu\hat{\phi}=-\sqrt{\frac{d-1}{d-2}}\varepsilon^\mu\nonumber\\
	&&{\cal S}^\nu \varepsilon^{\rho\sigma} = \frac{1}{2} \Bigl[\eta^{\sigma \nu}\,\eta^{\rho\mu}+\eta^{\rho \nu}\,\eta^{\sigma\mu}-\frac{2\eta^{\mu \nu}\eta^{\rho\sigma}}{d-2}\Bigl]\varepsilon_\mu(p_i)
\end{eqnarray}
and comparing these equations with equations (\ref{2.29ad}), \eqref{2.28} and \eqref{2.2.46}, we immediately see that they coincide if we identify ${\cal S}_\mu$ with the angular momentum operator $S^{z\rho}/\sqrt{2}$ associated to broken generators of the $d+1$ dimensional Lorentz group.

\subsection{Soft factorization with hard massive spin two  Kaluza-Klein states}
\label{Kaluza}

We have seen that the circle compactification of the graviton soft theorem gives rise to the soft factorization of amplitudes with soft scalar and gauge fields interacting with massless and massive states. When the finite energy states in $d+1$ dimension are gravitons, the Kaluza-Klein modes are massive spin two states charged with respect to the abelian gauge field. 
In this section, we shall consider these massive spin-2 KK modes interacting with a soft vector field and show that the explicit calculation of the amplitude agrees with the result obtained from compactification.

\vspace*{.07in} The action for the tower of massive spin two KK states interacting with an abelian gauge field should be obtained by considering the compactification of the Einstein-Hilbert action in the background determined by the massless modes of the metric. The lagrangian for the massive spin 2 KK modes, by the procedure of compactification, has been considered in \cite{cho&zho,cho&zho1} 
ignoring their interaction with the gauge field, (see also Ref.\cite{1910.04767} regarding the interaction with the massless fields). 
This gives the free Fierz-Pauli lagrangian \eqref{A.84la}.\footnote{The Fierz-Pauli lagrangian considered alone suffers from various problems (see, e.g., \cite{Rahman:2013sta}) as is usually the case with higher spin theories in flat space-times. However, the compactification procedure gives an infinite tower of  massive spin 2 fields that should 
form a consistent system. This can be seen for the case of compactification of 11 dimensional theory on $S^1$ since the resulting massive spin 2 tower in 10 dimension corresponds to D0 brane and its bound states.}  In the following, we shall use the Fierz-Pauli Lagrangian minimally coupled with an abelian gauge field (see for example \cite{Rahman}), reviewed in appendix \ref{apps1} (see equation \eqref{A.86}), for computing tree level amplitudes among the abelian gauge field and KK-states.  
We shall use the Feynman rules coming  from such an action to check the factorization properties obtained via compactification.\footnote{It should be noted that the Fierz-Pauli Lagrangian doesn't contain any cubic interaction term among massive fields of different KK levels and the massless fields. It turns out that even in the full KK action, obtained by compactification, such interaction terms can't arise. This is due to the fact that the charge conservation forbids cubic couplings between a massless gauge field and two massive particles of different KK-level.}

\vspace*{.07in}The 3-point vertex obtained from \eqref{A.86} giving the interaction among the vector with momentum $q$ and polarization $\varepsilon^\mu$ and two Kaluza-Klein states at the same level $n$, having momenta and polarizations $(k_2, \phi_{\mu \nu})$ and $(k_3,{\phi^*}_{\rho \sigma})$ is
\begin{eqnarray}
&&V_{\tau ;\rho \sigma;\mu \nu}(q,k_3,k_2)=\frac{i}{2} \hat{e} \Bigg[\frac{1}{2}(\eta_{\rho \mu}\eta_{\sigma\nu} +\eta_{\rho\nu}\eta_{\sigma \mu} -2 \eta_{\rho \sigma}\eta_{\nu\mu}) (k_2-k_3)_\tau+\frac{1}{2} \eta_{\tau \rho}\eta_{\mu \nu}(k_2-k_3)_\sigma\nonumber\\
&&+\frac{1}{2} \eta_{\tau \sigma} \eta_{\mu \nu} (k_2-k_3)_\rho+\frac{1}{2} \eta_{\tau \mu} \eta_{\rho \sigma} (k_2-k_3)_\nu +\frac{1}{2} \eta_{\tau \nu} \eta_{\rho \sigma} (k_2-k_3)_\mu-\frac{1}{2} \eta_{\tau \rho} (\eta_{\sigma\nu} k_{2\mu}+ \eta_{\mu \sigma} k_{2d\nu})\nonumber\\
&& -\frac{1}{2}\eta_{\tau \sigma}( \eta_{\rho \nu} k_{2\mu} +\eta_{\rho \mu} k_{2\nu})+\frac{1}{2} \eta_{\tau \mu} (k_{3 \rho}\eta_{\sigma\nu} +\eta_{\rho d} k_{3 \sigma}) +\frac{1}{2} \eta_{\tau \nu} (\eta_{\sigma\mu} k_{3 \rho} +\eta_{\rho\mu} k_{3\sigma} )\nonumber\\
&&+\frac{g}{2}
\eta_{\tau \mu} (q_{\sigma} \eta_{\rho\nu}+ q_{\rho} \eta_{\sigma\nu}) +\frac{g}{2} \eta_{\tau \nu} (q_{\sigma} \eta_{\rho\mu} +q_{\rho} \eta_{\sigma\mu})-\frac{g}{2} \eta_{\tau \sigma}(q_{\mu} \eta_{\rho \nu}+ q_{\nu} \eta_{\rho\mu}) -\frac{g}{2}\eta_{\tau \rho}(q_{\mu} \eta_{\sigma \nu} +q_{\nu} \eta_{\mu \sigma})\Bigg]\non\\
\label{4.65b}
\end{eqnarray}
The gyromagnetic ratio $g$, which appears in the above vertex, is a free parameter and is not fixed by the gauge invariance of the Lagrangian. It takes into account  the ambiguity in  the minimally coupled Fierz-Pauli lagrangian due to non-commutativity of the covariant derivatives\cite{Deser&Waldron}. 
\vspace*{.07in}The propagator of the massive states is given by  (see, e.g., \cite{Deser&Waldron})
\begin{eqnarray}
D^{\mu\nu\rho\sigma}= \frac{i}{p^2+m^2} \left[ \Pi^{\mu\rho}\Pi^{\nu\sigma} +\Pi^{\mu\sigma} \Pi^{\nu\rho}-\frac{2\,\Pi^{\mu\nu}\Pi^{\rho\sigma}}{d-1}\right]~~;~~\Pi^{\mu\nu}=\eta^{\mu\nu}+\frac{p^\mu p^\nu}{m^2}\label{4.66}
\end{eqnarray}
Here,  we have denoted with $\phi_{\mu\nu}$ and ${\phi^*}_{\mu\nu}$ the polarizations of the massive states.
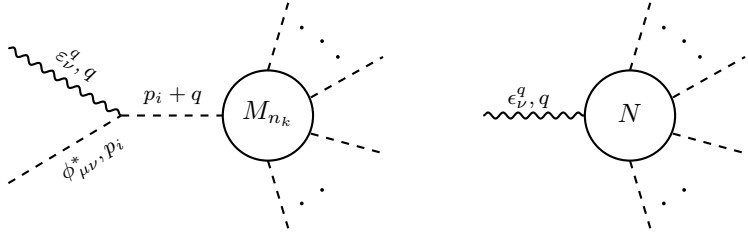
\begin{figure}
	\begin{center}
		\begin{tikzpicture}[scale=.30]
		\draw [thick, dashed]  (8.5,2) -- (9.4,5); 
		\draw [thick, dashed]  (10.4,0.8) -- (13.6,2.6); 
		\draw [thick, dashed]  (10.4,-0.8) -- (13.7,-1.7); 
		\draw [thick, dashed]  (8.5,-2) -- (9.4,-5); 
		\draw [thick, dashed] 
		(2,0)-- (6.5,0)  node[midway, sloped,above] {\scriptsize $\hspace*{.01in} p_i+q$};; 
		\draw [thick,decorate,decoration={snake,amplitude=.4mm,segment length=2mm,post length=0mm}] (-3,3) -- (2,0) node[midway, sloped,above] { \scriptsize $\hspace*{.01in}\varepsilon^q_\nu, q$};
		\draw[thick, dashed]  (-3,-3) -- (2,0) node[midway, sloped,below] { \scriptsize $\hspace*{.2in}{\phi^*}_{\!\!\!\mu\nu}, p_{i}$};
		\draw (8.5,.1) node { \footnotesize $M_{n_k}$};
		\draw [thick] (8.5,0) circle (2cm);
		\begin{scope}[shift={(-3,0)}] 
		\filldraw [ thick] (13.0,3.9) circle (1pt);
		\filldraw [ thick] (13.9,3.3) circle (1pt);
		\filldraw [ thick] (14.6,2.6) circle (1pt);
		\filldraw [ thick] (13.0,-3.9) circle (1pt);
		\filldraw [ thick] (13.9,-3.3) circle (1pt);
		\end{scope}
		\end{tikzpicture}
		\hspace*{1cm}
		\begin{tikzpicture}[scale=.30]
		\draw [thick, dashed]  (8.5,2) -- (9.4,5); 
		\draw [thick, dashed]  (10.4,0.8) -- (13.6,2.6); 
		\draw [thick, dashed]  (10.4,-0.8) -- (13.7,-1.7); 
		\draw [thick, dashed]  (8.5,-2) -- (9.4,-5); 
		\draw [thick,decorate,decoration={snake,amplitude=.4mm,segment length=2mm,post length=0mm}]        (2,0)   -- (6.5,0); 
		\draw (4,.8) node { \scriptsize $\epsilon_\nu^q, q$};
		\draw (8.5,.1) node { \footnotesize $N$};
		\draw [thick] (8.5,0) circle (2cm);
		\begin{scope}[shift={(-3,0)}] 
		\filldraw [ thick] (13.0,3.9) circle (1pt);
		\filldraw [ thick] (13.9,3.3) circle (1pt);
		\filldraw [ thick] (14.6,2.6) circle (1pt);
		\filldraw [ thick] (13.0,-3.9) circle (1pt);
		\filldraw [ thick] (13.9,-3.3) circle (1pt);
		\end{scope}
		\end{tikzpicture}
		\caption{Diagrams contributing to the interaction amplitude among a soft gauge field and $n$ Kaluza-Klein states.}    
	\end{center}
	\label{fig2ad}
\end{figure}
The on-shell amplitude with  a massless  vector field interacting with $n$ Kaluza-Klein modes 
is given by the sum of the diagrams having a pole in the massive modes and the contact terms, i.e. terms without  the propagator of  the massive particle (see figure 2) 
\begin{eqnarray}
M_{n+1}&\equiv& \varepsilon^\tau (M_{n+1})_{\tau}(q,\,p_1\dots, p_{n}) \non\\
&=&\varepsilon^\tau \, {\phi^*}^{\rho\sigma}\sum_{i=1}^{n}\left[V_{\tau ;\rho\sigma;\alpha\beta}(q,p_i,-p_ i-q)D^{\alpha\beta\mu\nu}(-p_i-q) M_{\mu\nu}(p_1,\dots p_i+q\dots,p_{n})\right]\non\\[.1cm]
&&+\ \varepsilon_{\mu}N_{n+1}^\mu(q,\,p_i,\,  p_{n}) 
\end{eqnarray}
By imposing the current conservation condition\cite{1406.6987}
\begin{eqnarray}
(p_i+q)^\mu \,M_{\mu\nu}(p_1,\dots p_i+q,\dots p_n)=(p_i+q)^\nu \,M_{\mu\nu}(p_1,\dots p_i+q,\dots p_n)=0,
\end{eqnarray}
we can replace, in the propagator, the symmetric tensor $\Pi_{\mu\nu}$ with the flat metric $\eta_{\mu\nu}$. Furthermore by imposing the on-shell conditions $p_i^\mu\phi^*_{\mu\nu}=p_i^\nu\phi^*_{\mu\nu}=0$, ${\phi^*}^\mu_{~\mu}=0$ and $M_{\mu\nu}=M_{\nu\mu}$,  the amplitude simplifies giving:
\begin{eqnarray}
M_{n+1}&=&\varepsilon_{\mu} \sum_{i=1}^{n} \hat{e}_i  \Bigg[\frac{p_{i}^\mu}{p_iq} +\frac{ g \,q_\nu}{2 p_i q} S_i^{\mu\nu} \Bigg] M_n(p_1,\dots p_i+q,\dots p_{n})
\nonumber\\
&&+\varepsilon_{\mu} \sum_{i=1}^{n} \hat{e}_i\,{\phi^*_i}^{\mu \nu}\frac{q_\nu\, \eta^{\rho_i\sigma_i}}{2(d-1)p_i  q} M_{\rho_i\sigma_i}(p_1,\dots p_i+q,\dots p_{n})+\varepsilon_{q\mu}\,N_{n+1}^{\mu}(q;\,p_1,\dots,  p_{n})\nonumber\\
&&\label{4.65}
\end{eqnarray}
The action of the spin operator $S_i^{\rho\sigma}$, given in equation \eqref{edefJ}, on the massive spin 2 states can be expressed as
\begin{eqnarray}
S_i^{\rho\sigma}{\phi_i^*}^{\mu\nu}= {\phi_i^*}^{\sigma \mu}(\Sigma^{\rho\sigma})_\sigma^{~\nu}+{\phi_i^*}^{\sigma \nu}(\Sigma^{\rho\sigma})_\sigma^{~\mu}.
\end{eqnarray}

The contact term $N^\mu$ can be fixed, in the infrared region where the momentum $q$ of the gauge field is small, by imposing the gauge invariance of the amplitude
\begin{eqnarray}
0&=&q^\mu M_\mu(q,\,p_1\dots, p_{n})\non\\
&=&M_n \sum_{i=1}^n \hat{e}_i +\sum_{i=1}^{n} \hat{e}_i\,q\cdot\frac{\partial}{\partial p_i} M_n(p_1,\dots p_i,\dots p_{n})+q_\mu N^\mu_{n+1}(0;p_1,\dots, p_{n})\nonumber\\
&& +\sum_{i=1}^{n} \frac{\hat{e}_i}{2}{\phi^*_i}^{\tau \delta}\frac{q_\tau\,q_\delta\, \eta^{\mu_i\nu_i}}{(d-1)p_i  q} M^{n}_{\mu_i\nu_i}(p_1,\dots p_{n})+O(q)
\end{eqnarray}  
By imposing the charge conservation $\sum_{i=1}^n e_i=0$ we get,
\begin{eqnarray}
&&\hspace*{-1.3cm}N^\mu_{n+1}(0;p_1,\dots, p_{n_k})\non\\
&=&-\sum_{i=1}^{n} \hat{e}_i\,\frac{\partial}{\partial p_{i\mu}} M_n(p_1,\dots p_i,\dots p_{n})-\sum_{i=1}^{n} \frac{\hat{e}_i}{2}{\phi^*_i}^{\mu \delta}\frac{\,q_\delta\, \eta^{\rho_i\sigma_i}}{(d-1)p_i  q} M_{\rho_i\sigma_i}(p_1,\dots p_{n})+ O(q)\nonumber\\ \label{Nn+1}
\end{eqnarray}
up to local terms of the form $E_\mu=(A\cdot q)B_\mu-(B\cdot q)A_\mu$\cite{1406.6987}.

\vspace*{.07in}By using \eqref{Nn+1} in \eqref{4.65}, we finally get
\begin{eqnarray}
M_{n+1}=\varepsilon_{\mu} \sum_{i=1}^{n} \hat{e}_i\,\Bigg[\frac{p_{i}^\mu}{p_i q} +\frac{q_\rho  }{2 p_iq}( L^{\mu\rho}_i+gS_i^{\mu\rho}) +\frac{ q_\rho}{  2p_iq}
L_i^{\mu\rho}\Bigg]M_n(p_i)+O(q)
\end{eqnarray}
This equation is consistent with \eqref{2.39intro} when we specialize that to the case of the massive spin 2 external states (in which case, last term in right hand side of \eqref{2.39intro} vanishes) provided we take the gyromagnetic ratio to be $g=1$.
We also notice  that $g=1/2$, rather than $g=1$,  is the gyromagnetic factor consistent with the counting of the degrees of freedom of the model \cite{Deser&Waldron}. However,
the full Kaluza-Klein theory that arises from the compactification of the Einstein-Hilbert action is not the free Fierz-Pauli action  but a theory with an  infinite number of degrees of freedom and it is not clear if the consistency conditions that fix $g$ to be $1/2$ are still applicable to our case. Indeed, a study of the connection between the Fierz-Pauli action minimally coupled to an abelian field and the Kaluza-Klein compactification of Einstein Hilbert action deserves a deeper analysis.

\section{Discussion}
\label{s4}
In this paper, we have analyzed the single soft graviton theorem under compactification. We considered the compactification on $S^1$ and showed that it gives a convenient tool for analyzing the soft behaviour of amplitudes in a particular dimension from the known soft behaviours in one higher dimension. 

\vspace*{.07in} However, this technique could also be useful in  considering compactification on more general manifolds such as hypertorus  or Calabi-Yau spaces. In these cases the spectrum that arises from the compactifications would be different and the technique presented in this paper could be useful in obtaining the soft behaviour of the particles which arise due to compactification on these manifolds.

\vspace*{.07in}In the case of the soft vector field, we have also shown explicitly (which is consistent with the result obtained from compactification) that the gauge invariance fixes the soft behaviour of the amplitudes, up to order $q^0$, even in presence of hard massive spin two fields. This check has not been possible to perform in the case of the soft scalar due to the absence of a symmetry which can determine the contact diagrams from the exchange diagrams. This has prevented the derivation of the soft scalar behaviour from some ward identity even though the scalar soft operators contain the generators of the scale transformations. In this respect, it is crucial to note that the soft scalar particle, arising from the circle compactification, is the  Goldstone boson associated with the breaking of a global scale invariance\cite{Duff1994} and it would be interesting to explore the connections between this breaking  and the soft scalar theorem as done in the case of the dilaton of spontaneously broken conformal field theories\cite{1507.08162,1512.03316}.

\vspace*{.07in}It would also be interesting to consider the interaction of soft RR 1-form field with the D0 branes and its bound states. The results should be identical to the case of external massive KK modes considered in section \ref{s3}. Usually, one considers the D branes as solitonic objects. However, treating them as external states participating in scattering from soft particles requires to treat them as perturbative objects. Moreover, we need to consider at least two D0 branes to get a non zero result since, otherwise, the result vanishes by conservation of charge. 

\vspace*{.07in}We have not considered the case of the multiple soft particles. However, the same technique can also be applied to this case. This should produce the multi soft behaviour of amplitudes in lower dimensions from the known multi particle soft theorems in higher dimensions \cite{Chakrabarti:2017ltl, AtulBhatkar:2018kfi}. In general, deriving the multi particle soft behaviour is more involved. However, using the tools of compactification, they can be easily obtained for a variety of particles.

\bigskip

{\bf Acknowledgement:} 
We thank Paolo Di Vecchia, Sitender Kashyap, Matin Mojaza, Wolfgang Mueck, Ashoke Sen and Massimo Taronna for useful discussions and for their helpful  comments on the manuscript.  We also thank the Galileo Galilei Institute for Theoretical Physics  for hospitality and partial support during the workshop "String Theory from a world-sheet perspective" where  this work started. {MV is also thankful to Harish Chandra Research Institute (HRI) for hospitality while this work was near completion. }

\appendix

\section{Review of compactification  on $S^1$}
\label{apps1}
In this appendix, we review some results about the dimensional reduction of the metric in  $(d+1)$ dimensions   on $S^1$. Gravity in $(d+1)$ dimensional  space-time  is described by the Einstein-Hilbert action
\begin{eqnarray}
S=\frac{1}{2\kappa_{d+1}^2} \int d^{d+1} x \sqrt{-G} \, R\label{app1e}
\end{eqnarray}
The $\kappa_{d+1}$ is related to the $(d+1)$-dimensional Newton's coupling constant as $2\kappa_{d+1}^2=16\pi G_N$. We parametrize the compact direction by $z$ and expand the metric in terms of its fourier modes on the circle as
\be
G_{MN}= \sum_{n=-\infty}^\infty G_{MN}^{(n)}(x)\ e^{\f{inz}{R_{d}}},\label{metric_expansion}
\ee
where $R_{d}$ is the radius of the compact direction.

\vspace*{.07in}The most general  compactification  ansatz  which is consistent with the diffeomorphism invariance in $d$ dimensions is 
\be
G_{\mu\nu} &=& e^{2\alpha\phi} g_{\mu\nu} \ +\ e^{2\beta\phi}A_\mu A_\nu\quad,\qquad G_{\mu z}  = e^{2\beta\phi} A_\mu\quad,\qquad G_{zz}= e^{2\beta\phi}\label{ansatzm2}
\ee
with the inverse metric and the determinant given by
\be
G^{\mu\nu} = e^{-2\alpha\phi} g^{\mu\nu} \quad,\qquad G^{\mu z}  &=& -e^{-2\alpha\phi} A^\mu \quad,\qquad G^{zz}= e^{-2\beta\phi}+e^{-2\alpha\phi}A_\mu A^\mu\label{ansatzm2}
\ee
\be
\mbox{det} (G_{\mu\nu})= e^{2(d\alpha+\beta)\phi }\ \mbox{det} (g_{\mu\nu})
\ee
$\alpha$ and $\beta$ are some arbitrary constants and the fields $g_{\mu\nu}(x,\,z)$, $\phi(x,\,z)$ and $A_\mu(x,\,z)$ depend on the full $d+1$ space-time coordinates. For the above metric ansatz, the Einstein-Hilbert action \eqref{app1e}, up to total derivative terms, takes the form (see e.g., \cite{cho&zho})
\begin{eqnarray}
S&=&\frac{1}{2\kappa_{d+1}^2} \int d^dx \int_0^{2\pi R_d} dz \sqrt{g}\Bigg\{ e^{(\beta+(d-2)\alpha)\phi} R-\frac{1}{4} e^{((d-4)\alpha+3\beta)\phi} F^{\mu\nu}F_{\mu\nu}\non\\
&&+\bigl[2\alpha (d-1) (\beta+(d-2)\alpha) -\alpha^2(d-2)(d-1)\bigl]e^{(\beta +(d-2)\alpha)\phi}\partial_\mu\phi\partial^\mu\phi \nonumber\\
&&+\frac{1}{4} e^{((d-2)\alpha-\beta)\phi} g^{\mu\nu}g^{\rho\sigma}\big[\partial_z (e^{2\alpha \phi} g_{\mu\rho})\partial_z(e^{2\alpha \phi} g_{\mu\sigma}) -\partial_z (e^{2\alpha \phi}g_{\mu\nu})\partial_z(g_{\rho\sigma}e^{2\alpha \phi})\big]\Bigg\}\label{5.2.6}
\end{eqnarray}
where $F_{\mu\nu}$ denotes the field strength of the vector field $A_\mu$.

We now focus on the zero modes in the KK expansion \eqref{metric_expansion}. These zero modes do not depend on the compact coordinate and represent the massless degrees of freedom in the $d$-dimensional theory. More precisely, these zero modes describe the metric, a gauge field and a scalar field in $d$ dimensions. There are some specific choices for the constants $\alpha$ and $\beta$ for the zero modes. E.g., if we want to obtain the dimensionally reduced action in the Einstein frame with the canonically normalized scalar kinetic term, we need to choose 
\be
\beta=(2-d)\alpha\qquad, \qquad \alpha^2=\f{1}{2(d-1)(d-2)}
\ee
with these choices, the action for the zero modes of the metric reduces to (dropping the zero index from the fields and taking the negative root for $\alpha$)
\be
S_0&=&\f{1}{2\kappa^2_d}\int d^dx\sqrt{-g}\biggl[R_g-\f{1}{2}\p_\mu \phi\p^\mu \phi -\f{1}{4}e^{\sqrt{2(d-1)/(d-2)}\, \phi}\ F_{\mu\nu}F^{\mu\nu}\biggl] \label{dim_red}
\ee
where we defined $\kappa_d^2=\f{\kappa_{d+1}^2}{2\pi R_d}$. 

\vspace*{.07in}Similarly, for going to the string frame, we need to choose
\be
\alpha\ =\ \f{3}{1-d}\qquad,\qquad \beta\ =\ \f{d-4}{d-1}
\ee
With these choices, for $d+1=11$, the action for the zero modes in $d=10$ reduces to
\be
S_0&=&\f{1}{2\kappa^2_{10}}\int d^{10}x\sqrt{-g}\ \biggl[e^{-2\phi}\left(R_g+4\p_\mu \phi\p^\mu \phi\right)-\f{1}{4}F_{\mu\nu}F^{\mu\nu} \biggl] \label{dim_red1}
\ee
Next, we consider the non-zero modes. Their analysis is more involved and is carried out in some detail, for example  in \cite{cho&zho,cho&zho1} for $d+1=5$ and for the free theory neglecting the interaction of the KK-modes with  the massless fields graviton, vector and scalar. Here, we follow a slightly different approach which is closer to reference\cite{Nappi&Witten}.  Assuming $d$-dimensional Poincar\'e invariance of the vacuum, we impose the conditions\cite{Duff&Dolan}
\begin{eqnarray}
\langle g_{\mu\nu}\rangle =\eta_{\mu\nu} ~~;~~ \langle A_\mu\rangle =0~~;~~\langle e^{\phi}
\rangle =1
\end{eqnarray}
and expand  the metric around such a background as
\begin{eqnarray}
G_{\mu\nu}=\eta_{\mu\nu} +2\kappa_{d+1} S_{\mu\nu}(x,\,z)~~,~~G_{zz}=1+2\kappa_{d+1}S_{zz} ~~;~~G_{\mu z}=2\kappa_{d+1}S_{\mu z}  \label{expd+1}
\end{eqnarray}
The KK expansion for the non zero modes is given by
\begin{eqnarray}
\tilde{S}_{\mu\nu}=\sum_{n\neq 0}S_{\mu\nu}^{(n)}\, e^{i p_z z} ~~;~~\tilde{S}_{\mu z}=\sum_{n\neq 0}S_{\mu z}^{(n)}\, e^{i p_z z}~~;~~\tilde{S}_{zz}=\sum_{n\neq 0}S_{zz}^{(n)}\, e^{ip_z z}\label{A.84}
 \end{eqnarray}
where $\tilde{S}$ denotes the non zero modes of the KK-expansion of the metric, $z\in[0,\,2\pi R_d]$ and $p_z=n/R_d$.

\vspace*{.07in}The $d+1$ dimensional parametrization invariance of the theory allows us to gauge fix the fields $\tilde{S}_{\mu z}$ and $\tilde{S}_{zz}$ to zero \cite{Nappi&Witten}. According to equation \eqref{ansatzm2}, this corresponds to fixing the  non zero modes of the scalar and gauge fields  to zero. We can gauge away these fields because they act as Goldstone fields and   $S_{\mu\nu}^{(n)}$  eats them to become a massive spin 2 particle,  as we are going to see.
  
\vspace*{.07in}Equation \eqref{5.2.6}, restricted only to the non zero modes of the fields, simplifies and at lowest order in the field expansion becomes
\begin{eqnarray}
S^{n.z.m}&=& \int d^d x \int_0^{2\pi R_d}d z\bigg[\frac{1}{2} \partial_\rho \tilde{S}_{\mu\nu}\partial^\rho \tilde{S}^{\mu\nu}-\frac{1}{2}\partial_\rho \tilde{S} \partial^\rho \tilde{S}+ \partial_\mu \tilde{S}\partial_\nu \tilde{S}^{\mu\nu} -\partial_\rho \tilde{S}_{\mu\nu} \partial^\nu \tilde{S}^{\mu\rho}\nonumber\\
&&+\,\frac{1}{4} \bigg( \partial_z \tilde{S}_{\mu\nu} \partial_z\tilde{S}^{\mu\nu}-\partial_z \tilde{S}\partial^z\tilde{S}\bigg)\Bigg] \label{A.85}
\end{eqnarray} 
with $\tilde{S}=\tilde{S}^\mu_{~\mu}$.  

\vspace*{.07in}By inserting in equation \eqref{A.85} the mode expansion given in \eqref{A.84} and introducing the $d$ dimensional fields $\phi^{(n)}_{\mu\nu} =\sqrt{2\pi R_d} \,S^{(n)}_{\mu\nu}$
we get, for each level $n$ of the Kaluza-Klein mode expansion, the Fierz-Pauli lagrangian\cite{Fierz&Pauli}
\begin{eqnarray}
{\cal L}&=&\frac{1}{2}\partial_\mu\phi^{(-n)}_{\nu\rho}\partial^\mu{\phi^{(n)}}^{\nu\rho} 
-\partial_\mu{\phi^{(-n)}}^{\mu\nu}\partial^\rho\phi^{(n)}_{\rho\nu} -\frac{1}{2} \partial_\mu{\phi^{(-n)}}\partial^\mu\phi^{(n)}+\frac{1}{2}\partial_\mu{\phi^{(-n)}}^{\mu\nu}\partial_\nu {\phi^{(n)}}\nonumber\\
&& +\frac{1}{2}\partial_\mu{\phi^{(n)}}^{\mu\nu}\partial_\nu {\phi^{(-n)}}+\frac{m_n^2}{2} (\phi_{\mu\nu}^{(-n)}{\phi^{(n)}}^{\mu\nu} -{\phi^{(-n)}}\phi^{(n)})\label{A.84la}
\end{eqnarray}
with $\phi=\phi^\mu_{~\mu}$,  $\phi^{(n)}_{\mu\nu}=(\phi^{(-n)}_{\mu\nu})^*$ and   $m_n^2 =\frac{n^2}{R_d^2}$.

\vspace*{.07in}The complex fields $\phi^{(n)}_{\mu\nu}$ satisfy the equations of motion of a massive particle with mass $m_n^2=n^2/R_d^2$, i.e. (see for example \cite{Deser&Waldron})
\be
\Bigl(\Box+\f{n^2}{R_d^2}\Bigl)\phi_{\mu\nu}^{(n)}(x)\ =\ 0~~;~~\partial^\mu\phi_{\mu\nu}^{(n)}=\partial^\nu\phi_{\mu\nu}^{(n)}={\phi^{(n)}}^\mu_{~\mu}=0
\ee   
These are an infinite tower of massive modes with masses given by
\begin{eqnarray}
m_n^2=-p_d^2=-p_z^2=\frac{n^2}{R_d^2}
\end{eqnarray}
Thus, the $d$ dimensional compactified theory has an infinite tower of the massive Kaluza Klein states.
These massive KK modes are also charged with respect to the massless $U(1)$ gauge field $A_\mu^{(0)}$.
This happens because the zero mode of the diffeomorphism along the compact direction, namely, $\delta x^z=-\xi^z(x^\mu)$, becomes a local gauge transformation for the $d$-dimensional vector field, $A^{(0)}_\mu\rightarrow A^{(0)}_\mu+ \partial_\mu \xi^z(x^\mu)$\cite{Duff&Dolan}. From equation \eqref{A.84}, we easily see that the massive modes transform under such transformation as $S_{\mu\nu}^{(n)}\rightarrow S_{\mu\nu}^{(n)} e^ {-i p_z  \xi^z}$ and therefore carry  the charge $e\equiv  p_z$ with respect to this $U(1)$ group. The charge with respect to the canonically normalized field defined in \eqref{2.11} turns out to be $\hat{e}_n=\sqrt{2}\kappa_d p_z$\cite{Duff1994}. Below, we work with the canonically normalized field.

\vspace*{.07in}In this paper, in order to check the soft theorem statement involving a soft vector interacting with the tower of KK-states we need to include in equation \eqref{A.84la}, the interaction terms involving the gauge field. 
This is easily achieved by replacing the normal derivatives by the gauge covariant derivative, namely,
$\partial_\mu\rightarrow D_\mu= \partial_\mu +i\hat e_n \hat A_\mu$. Since the covariant derivatives do not compute, the minimal coupling procedure is ambiguous and this ambiguity is parametrized  by a constant $g$ which is called the gyromagnetic ratio. Thus, the Fierz-Pauli Lagrangian which includes the interaction with an abelian gauge field turns out to be (see for example \cite{Deser&Waldron})
\begin{eqnarray}
{\cal L}&=&\frac{1}{2}D_\mu\phi^*_{\nu\rho}D^\mu{\phi}^{\nu\rho} 
-D_\mu{\phi^*}^{\mu\nu}D^\rho\phi_{\rho\nu} -\frac{1}{2} D_\mu{\phi^*}D^\mu\phi+\frac{1}{2}D_\mu{\phi^*}^{\mu\nu}D_\nu {\phi}\nonumber\\
&& +\frac{1}{2}D_\mu{\phi^*}^{\mu\nu}D_\nu {\phi}+\frac{m_n^2}{2} (\phi^*_{\mu\nu}{\phi}^{\mu\nu} -{\phi^*}\phi)-i \hat e g {\phi^*}^{\rho\mu}F_{\mu\nu}\phi^\nu_{~\rho}\label{A.86}
\end{eqnarray}
with $F_{\mu\nu}=\partial_\mu\hat{A}_\nu -\partial_\nu \hat{A}_\mu$ and we have omitted the label $n$ and used $\phi^{(-n)}=\phi^*$  for having  a lighter notation.  Equation \eqref{A.86} is the starting point for getting the Feynman rules written in equations \eqref{4.65b} and \eqref{4.66}.
 
\vspace*{.07in}For the compactification of 11 dimensional theory on $S^1$, the massive KK modes form the short 256-dimensional susy multiplets and hence they are all BPS states. As mentioned above, the mass (or equivalently conserved U(1) charge) of these states is given by $m=|n|/R_d$. It turns out that in type IIA string theory in 10 dimensions, there are objects with precisely the same properties, namely D0 branes. They also form the short 256-dimensional representation of susy algebra. The tension (or mass) of the D0 branes is given by $1/(g_s\sqrt{\alpha'})$. Hence, being the BPS states, they also carry the U(1) charge in multiples of $1/(g_s\sqrt{\alpha'})$. This means that a single D0 brane can be identified with the $n=1$ KK modes for the radius of compactification $R_{10}=g_s\sqrt{\alpha'}$. The higher KK modes are then identified with the bound states of D0 branes.

\end{document}